\documentclass[11pt,a4paper]{article}

\usepackage[T1]{fontenc}
\usepackage{lmodern,microtype}
\usepackage[margin=25mm]{geometry}
\usepackage{amsmath,amssymb,amsthm,mathrsfs,tikz-cd}
\numberwithin{equation}{section}
\newtheorem{theorem}{Theorem}[section]
\newtheorem{proposition}[theorem]{Proposition}
\newtheorem{lemma}[theorem]{Lemma}
\theoremstyle{remark}
\newtheorem{remark}[theorem]{Remark}

\usepackage{booktabs,enumitem,xcolor,float}
\usepackage{cite}
\setlist{itemsep=3pt,topsep=5pt}

\usepackage{hyperref}
\hypersetup{colorlinks=true,linkcolor=blue!45!black,citecolor=blue!45!black,
 urlcolor=blue!55!black,pdftitle={From pure spinor modules to superspace superfields},
 pdfauthor={Ond\v{r}ej Hul\'ik and Raphael Senghaas}}

\newcommand{\C}{\mathbb C}
\newcommand{\A}{A^\bullet}
\newcommand{\U}{C^\infty(T)}
\newcommand{\Ch}{\mathcal C}
\newcommand{\calO}{\mathcal{O}}
\newcommand{\OY}{{\mathcal O_Y}}
\newcommand{\OYd}[1]{{\mathcal O_{Y_{#1}}}}
\newcommand{\CE}[1]{\mathrm{CE}(\mathfrak n_{#1})}
\newcommand{\G}{\mathcal G}
\newcommand{\Pot}{\mathcal P}
\newcommand{\bD}{\bar D}
\newcommand{\Cone}{\operatorname{Cone}}
\newcommand{\Ext}{\operatorname{Ext}}
\newcommand{\Tor}{\operatorname{Tor}}
\newcommand{\Hom}{\operatorname{Hom}}
\newcommand{\RHom}{\operatorname{RHom}}
\newcommand{\Ann}{\operatorname{Ann}}
\newcommand{\str}{\mathfrak s}
\newcommand{\nodd}{\mathfrak n_{\mathrm{odd}}}

\newcommand{\lp}{\lambda_+}
\newcommand{\lm}{\lambda_-}
\newcommand{\lbp}{\bar\lambda_+}
\newcommand{\lbm}{\bar\lambda_-}
\newcommand{\bpl}{{\bar{+}}}
\newcommand{\bmi}{{\bar{-}}}
\newcommand{\pplus}{{+\!\!\!+}}
\newcommand{\mminus}{{=}}
\newcommand{\pmpm}{{\stackrel{+}{=}}}
\newcommand{\vpp}{v_{\pplus}}
\newcommand{\vmm}{v_{\mminus}}

\title{From pure spinor modules to superspace superfields}
\author{Ond\v{r}ej Hul\'ik$^{a,b}$ \quad and \quad Raphael Senghaas$^c$\\[6pt]
{\small\normalfont $^a$Institute for Mathematics, Ruprecht-Karls-Universit\"at Heidelberg,}\\
{\small\normalfont 69120 Heidelberg, Germany}\\
{\small\normalfont $^b$Department of Theoretical Physics and Astrophysics, Faculty of Science, Masaryk University,}\\
{\small\normalfont Kotl\'a\v{r}sk\'a 2, Brno, Czechia}\\
{\small\normalfont $^c$Beijing Institute of Mathematical Sciences and Applications (BIMSA),}\\
{\small\normalfont Huairou District, Beijing 101408, China}\\[6pt]
{\small\normalfont\href{mailto:ondra.hulik@gmail.com}{\texttt{ondra.hulik@gmail.com}},\quad
\href{mailto:raphaelsenghaas@bimsa.cn}{\texttt{raphaelsenghaas@bimsa.cn}}}}
\date{}

\begin{document}
\maketitle

\begin{abstract}
	We relate the pure spinor formalism to the superspace description of
	constrained superfields and gauge multiplets, mainly for abelian
	two-dimensional $\mathcal N=(2,2)$ supersymmetry, with three- and
	four-dimensional $\mathcal N=1$ examples for comparison. For each multiplet
	we give an explicit module over the supertranslation Chevalley--Eilenberg
	algebra and match its reconstruction with the known superfields.
	First-order constraints come from modules supported on linear subspaces of
	the nilpotence variety, and higher-order ones, such as the complex-linear
	constraint, from Yoneda products of their Ext classes. Requiring a
	connection to act on given matter and its conjugate restricts its
	curvature, which yields the ordinary, twisted and semichiral vector
	multiplets with their field strengths. The module of each vector multiplet
	is the fiber product over the origin of the coordinate rings of the matter
	supports: its potentials are differences of partial potentials, and its
	gauge parameters are the matter superfields. For the large vector
	multiplet this fiber product is not a quotient of the pure spinor ring;
	its field strengths are four odd semichiral superfields, and partial
	regluing gives its constrained versions.
\end{abstract}
\begingroup\small\setlength{\parskip}{0pt}\tableofcontents\endgroup

\bigskip

\section{Introduction}

How strongly supersymmetry constrains a model depends largely on whether it
can be realized off shell.
Where this is possible, superspace gives manifestly supersymmetric
actions and a direct route to quantization and quantum computations through
supergraphs \cite{Superspace}. Multiplets appear as superfields, i.e.\ functions
or sections on superspace, usually subject to differential constraints and
gauge transformations. Such formulations, however, exist only for a limited
range of dimensions and amounts of supersymmetry.

An alternative that has attracted renewed attention is the pure spinor
superfield formalism \cite{Berkovits,CNT,CederwallOverview}, which works in
every dimension and for any amount of supersymmetry. A multiplet is encoded
as a module over the Chevalley--Eilenberg (CE) algebra of the
supertranslation algebra. Geometrically, the cohomology of
the CE algebra in lowest translation-ghost degree is the coordinate ring of the nilpotence
variety, and modules over it play the role of sheaves on this variety
\cite{Nilpotence,Perspectives}. By the derived equivalence of Elliott, Hahner
and Saberi, every multiplet in the sense of \cite{Derived} arises from some
module \cite[Theorem~4.3]{Derived}.

One drawback of the formalism is that the module does not tell us in
advance what kind of multiplet it produces. The result may be a fully
off-shell multiplet (as we will see in Section~\ref{sec:matter}) or a BV
multiplet, in which the equations of motion are built in; in ten
dimensions the pure spinor ring gives the BV multiplet of super Yang--Mills
theory \cite{CNT,CederwallOverview}, for which no off-shell formulation with
finitely many auxiliary fields is known. Where a superspace formulation
exists, suitable modules reproduce the off-shell component multiplets,
including auxiliary fields such as $F$ and $D$
\cite[Sections~3.5 and~3.8]{Perspectives}.

The relation between the two languages is often left implicit: given a
module, which constrained superfield does it describe, and given a
constraint, which module produces it? When an off-shell superspace
formulation exists, answering these questions combines the generality of the
pure spinor formalism with the practical strength of superspace. Here we
answer them for the low-dimensional multiplets we consider. The passage from
modules to superfields is an explicit computation; the passage back is guided
by the inverse functor of \cite[Theorem~4.3]{Derived} and checked case by
case:
\begin{equation}
	\text{module and morphism}\quad\longleftrightarrow\quad
	\text{superfields, constraints, and gauge equivalences}.
	\label{eq:dictionary}
\end{equation}

In this paper we focus mainly on two-dimensional $\mathcal N=(2,2)$
supersymmetry, which has more kinds of constrained matter (chiral,
twisted-chiral and semichiral) and correspondingly more vector multiplets
than four-dimensional $\mathcal N=1$. We establish the dictionary for chiral,
twisted-chiral and semichiral matter \cite{GHR,BLR,LRvUZ}, for
complex-linear and twisted complex-linear superfields, and for the
ordinary, twisted, semichiral and large vector multiplets
\cite{NewVectors,GatesMerrell,Nonabelian,Ryb}. All of these are off-shell
multiplets. Alongside them we discuss examples from three-dimensional
$\mathcal N=1$ and four-dimensional $\mathcal N=1$ supersymmetry, to
illustrate which structures generalize and which are specific to two
dimensions. A pure spinor description of the four-dimensional vector
multiplet, whose dimensional reduction is the ordinary two-dimensional
vector, already appears in \cite[Section~3.8]{Perspectives}; extended supersymmetric quantum
mechanics is treated in \cite{ENSW}; six-dimensional multiplets built from
vector bundles, together with general results on short exact sequences and
dualities, in \cite{HNSW}; and duals of multiplets in \cite{Canonical} and
\cite[Corollary~4.9]{Derived}. The twisted versions of
the ordinary constructions follow by partial conjugation, the exchange
$\lm\leftrightarrow\lbm$. The physical multiplets, with their field
strengths, potentials and gauge transformations, are known from superspace,
and our results concern their module-side origin. In particular, to our knowledge
the derived description of complex-linear matter and the gauge complexes of
the semichiral and large vector multiplets have not been given before.

\subsection{Main results}
\begin{itemize}
	\item \emph{Second-order constraints are derived morphisms.} There is no module
	      map, and no class of degree one, from the augmentation module $k$ to the
	      chiral module $M_\Phi=\C[\lbp,\lbm]$. The first class appears in degree two; in the relevant
	      weight it is unique (Appendix~\ref{app:linear-class}); and its reconstruction is
	      $-\bD_+\bD_-$, so that its cone is the complex-linear multiplet shifted by one
	      (see \eqref{eq:MSigma}). More generally, for nested modules of linear support the
	      Yoneda product of $m$ first-order classes reconstructs to a product of $m$
	      spinor derivatives (Proposition~\ref{prop:yoneda}): the higher-order
	      constraint is a first-order system with auxiliary fields, and the Yoneda
	      product eliminates them. The class of complex-linear matter is the case
	      $m=2$, and it extends to four dimensions over the pure spinor ring $\OYd{4}$.
	\item \emph{Further presentations of complex-linear matter.} In two dimensions
	      the complex-linear module is realized by a finite strict CE-module on which
	      the translation ghosts act (Appendix~\ref{app:finite-model}), and, up to
	      shift, it is the Gorenstein dual of the augmentation ideal of the chiral
	      module (Theorem~\ref{thm:dual}).
	\item \emph{Matter constrains curvature.} Requiring one connection to act on
	      prescribed matter and its conjugate determines which quadratic curvature
	      components may be nonzero (Proposition~\ref{prop:integrability}).
	      Taking the largest allowed set among conjugation-invariant monomial
	      choices, chiral matter selects the ordinary vector ring, twisted-chiral
	      matter the twisted vector ring and the four kinds of semichiral matter
	      the semichiral vector ring, while a connection acting on chiral and
	      twisted-chiral matter and their conjugates is flat (see
	      \eqref{eq:flatness}).
	\item \emph{Allowed curvature gives the field strengths of the multiplet.} For each
	      selected ring the ideal of allowed curvature components reconstructs exactly the space
	      of field strengths (Proposition~\ref{prop:strengths}). In particular the chiral and
	      twisted-chiral strengths of the semichiral vector are independent, and
	      locally its gauge-invariant content is that of an ordinary and a twisted
	      vector multiplet together (see \eqref{eq:mv}). The same principle covers
	      complex-linear matter: its module has a two-term model by modules of
	      linear support, and twisting it by a connection gives the covariant
	      complex-linear multiplet exactly under the barred-curvature conditions
	      of charged chiral matter (Section~\ref{sec:selection}).
	\item \emph{Gauge complexes.} Each vector multiplet is the mapping cone of
	      its gauge transformation, and we obtain this cone directly from the module
	      (Section~\ref{sec:cones}). For the $3d$ vector this is immediate. For the
	      $4d$ vector and all $2d$ vectors, the module of the multiplet is the
	      fiber product over the origin of the polynomial rings on the supports of
	      the matter it gauges, a few coordinate planes or axes; the familiar
	      potentials arise as differences of potentials defined on these pieces,
	      and the gauge parameters are the chiral, twisted-chiral or semichiral
	      superfields belonging to each piece. Setting one field strength
	      of the semichiral vector to zero, which reduces it to the ordinary or the
	      twisted vector, is the reconstruction of a surjection of rings
	      (Section~\ref{sec:reductions}). For the large vector multiplet the
	      pieces are the four planes of chiral and twisted-chiral matter, which
	      overlap along axes, and the fiber product is not a quotient of the pure
	      spinor ring (see \eqref{eq:lvm-complex}); its gauge-invariant content is exactly four odd
	      semichiral field strengths (Theorem~\ref{thm:lvm-strengths}). Regluing
	      the planes only along some of the axes sets the corresponding strengths to zero and
	      gives constrained large vector multiplets, among them the multiplet of
	      \cite{BKK} (Proposition~\ref{prop:clvm}).
\end{itemize}

\paragraph{Organization.}
Section~\ref{sec:conventions} sets up the CE algebras, the reconstruction
functor, and the cone through which gauge redundancy enters.
Section~\ref{sec:matter} treats first-order constraints from modules of linear support,
spinor derivatives as derived morphisms, and their application to
complex-linear matter.
Section~\ref{sec:gauge} shows how matter constrains curvature and how the
allowed curvature components become field strengths.
Section~\ref{sec:vectors} solves the curvature constraints by partial
potentials and identifies the gauge complexes of all vector multiplets,
including the reductions of the semichiral vector and the large vector
multiplet. Section~\ref{sec:duality} discusses Gorenstein duality and
Section~\ref{sec:conclusions} concludes; the appendices give the complex-linear
Ext calculation and its finite CE model, and Betti numbers.

\section{From modules to superfields}\label{sec:conventions}

\subsection{CE algebras}\label{sec:setting}
In this section we recall the CE algebra of the supertranslation algebra,
its modules, and how multiplets are obtained from modules, following
\cite{Derived,Perspectives}.
When we say that a module reconstructs a known
multiplet, we mean a local quasi-isomorphism between the complex of superfields obtained from
the module as in \cite{Derived} and the superspace complex of constrained
superfields, or of potentials with their gauge transformations. Here and
below, a \emph{local quasi-isomorphism} is a morphism of complexes of sheaves
on superspace that induces isomorphisms on the cohomology of all stalks, and
``locally'' always refers to this notion.
In proofs we sometimes use auxiliary constructions that break
supersymmetry, such as contracting homotopies, explicit preimages built with
$\theta$ or $\bar\theta$, and primitives from the odd Poincar\'e lemma.
They only serve to compute cohomology; the quasi-isomorphisms we establish
are themselves supersymmetric.

The running examples we consider are $d=2$ with $\mathcal N=(2,2)$ supersymmetry and $d=3,4$ with
$\mathcal N=1$. For the corresponding supertranslation algebra
$\mathfrak n_d$\footnote{The subscript records only the dimension. Unlike
	$\mathfrak n_3$ and $\mathfrak n_4$, which are the minimal $\mathcal N=1$
	algebras, $\mathfrak n_2$ is not the minimal two-dimensional algebra but the
	$\mathcal N=(2,2)$ one, with four real supercharges, the dimensional
	reduction of four-dimensional $\mathcal N=1$.} we write $\CE{d}$ for its CE algebra, denoted
$C^\bullet(\mathfrak n)$ in \cite{Derived}; it is the BRST complex of the
supersymmetry algebra. Its \emph{spinor ghosts} $\lambda$, dual to the
supercharges, commute and have weight one; its \emph{translation ghosts} $v$,
dual to the translations, anticommute and have weight two. The CE algebra
fixes the torsion of superspace; it does not choose a multiplet or a
reality structure, and it does not fix dynamics. Up to the normalization of $v$,
\begin{align}
	\CE{3}                & =\C[\lambda^1,\lambda^2]\otimes\Lambda(v^{11},v^{12},v^{22}),                     &
	dv^{\alpha\beta}      & =\lambda^\alpha\lambda^\beta,                                                       \\
	\CE{4}                & =\C[\lambda^\alpha,\bar\lambda^{\dot\alpha}]\otimes\Lambda(v^{\alpha\dot\alpha}), &
	dv^{\alpha\dot\alpha} & =\lambda^\alpha\bar\lambda^{\dot\alpha},                                            \\
	\CE{2}                & =\C[\lp,\lbp,\lm,\lbm]\otimes\Lambda(\vpp,\vmm),                                  &
	d\vpp                 & =\lp\lbp,\quad d\vmm=\lm\lbm.
\end{align}
We call the quadrics $dv$ the \emph{torsion quadrics}.
Their cohomology in the lowest translation-ghost degree is the ring of
functions on the nilpotence variety \cite{Nilpotence,Perspectives}. Let
$Y_d$ denote the nilpotence variety, the scheme of square-zero odd elements
of $\mathfrak n_d$, and $\OYd{d}$ its structure sheaf, which we identify
with its graded ring of global sections, the pure spinor ring; in two
dimensions we write simply $Y$ and $\OY$. Explicitly,
\[
	\OYd{3}=\C[\lambda^1,\lambda^2]/(\lambda^1,\lambda^2)^2,\quad
	\OYd{4}=\C[\lambda^\alpha,\bar\lambda^{\dot\alpha}]/(\lambda^\alpha\bar\lambda^{\dot\alpha}),\quad
	\OY=R/I_T,
\]
where in two dimensions
\[
	R=\C[\lp,\lbp,\lm,\lbm],\qquad I_T=(\lp\lbp,\,\lm\lbm).
\]
We call $I_T$ the \emph{torsion ideal}: its generators are the images of the translation ghosts under the CE
differential, $\lp\lbp=d\vpp$ and $\lm\lbm=d\vmm$.
Geometrically $\OYd{3}$ is a
nonreduced point, $\OYd{4}$ is the union of two planes meeting at the origin,
and $\OY$ is the union of the four coordinate planes spanned by one of
$\lp,\lbp$ and one of $\lm,\lbm$. In three dimensions the scheme structure
is essential: the reduced point would not give the vector multiplet.

In two dimensions the map $\CE{2}\to\OY$ is a quasi-isomorphism. Indeed,
$\CE{2}$ is the Koszul complex over $R$ of $\lp\lbp$ and $\lm\lbm$, and these
form a regular sequence, being monomials in disjoint sets of variables.

In three and four dimensions this fails: $\CE{d}$ has cohomology beyond
$\OYd{d}$. The elements
\[
	\zeta^\alpha=\lambda_\beta v^{\alpha\beta}\in\CE{3},\qquad
	\varepsilon_{\alpha\beta}\lambda^\alpha v^{\beta\dot\gamma},\ \
	\varepsilon_{\dot\alpha\dot\beta}\bar\lambda^{\dot\alpha}v^{\gamma\dot\beta}\in\CE{4},
\]
with $\lambda_\beta=\varepsilon_{\beta\gamma}\lambda^\gamma$ and
$\varepsilon_{12}=\varepsilon_{\dot1\dot2}=1$ (for instance
$\zeta^1=\lambda^2v^{11}-\lambda^1v^{12}$ and
$\lambda^1v^{2\dot1}-\lambda^2v^{1\dot1}$), have weight three and are
closed. They are not exact: $d$ removes one translation ghost, so an exact
element with one translation ghost is the image of an element with two,
which has weight at least four. These elements span the weight-three
cohomology with one translation ghost, two classes in three dimensions and $2+2$ in four, and they
carry the field strengths of the three- and four-dimensional vector
multiplets (Section~\ref{sec:curv-ce}).

For this reason we always work with $\CE{d}$-modules and never replace
$\CE{3}$ or $\CE{4}$ by the ring of
functions on the nilpotence variety. The modules we use in three
and four dimensions are $\OYd{d}$-modules, regarded as $\CE{d}$-modules
through the map $\CE{d}\to\OYd{d}$. A class computed over $\OYd{d}$, as for
the four-dimensional complex-linear constraint, then gives a class
over $\CE{d}$, but uniqueness statements refer to $\OYd{d}$ only.

Even in two dimensions, where $\CE{2}\simeq\OY$ and nothing is lost by
working over $\OY$, modules on which the translation ghosts act
nontrivially are useful: they give the complex-linear superfield a finite,
strict model (Appendix~\ref{app:finite-model}).

We use standard superspace conventions. In two dimensions we write
$\lambda_i$, $i\in\{+,\bpl,-,\bmi\}$, for $(\lp,\lbp,\lm,\lbm)$ and $D_i$
for the corresponding spinor derivatives $(D_+,\bD_+,D_-,\bD_-)$, whose
only nonzero brackets are
\begin{equation}
	\{D_\pm,\bD_\pm\}=2i\partial_{\pmpm},\label{eq:derivatives}
\end{equation}
with $\partial_{\pplus},\partial_{\mminus}$ the light-cone derivatives and
$\pmpm$ matching the sign on the left. Two
ghosts $\lambda_i,\lambda_j$ form a \emph{torsion pair} if
$\{D_i,D_j\}\neq0$, that is, if $\{i,j\}$ is $\{+,\bpl\}$ or $\{-,\bmi\}$.
In four and three dimensions we use
$\{D_\alpha,\bD_{\dot\alpha}\}=-2i\partial_{\alpha\dot\alpha}$ and
$\{D_\alpha,D_\beta\}=2i\partial_{\alpha\beta}$.

\subsection{The reconstruction functor and component fields}
Let $T=\exp(\mathfrak n_d)$ be the supertranslation group. As a
supermanifold it is superspace, with even coordinates $x$ and odd
coordinates $\theta$; in two dimensions $T\cong\mathbb R^{2|4}$, with
coordinates $x^{\pplus}$, $x^{\mminus}$, $\theta^\pm$ and $\bar\theta^\pm$. We write $C^\infty(T)$ for the
sheaf of smooth complex-valued functions on $T$, which we identify with the
space of unconstrained
superfields, and $C^\infty_{\mathbb R}(T)$ for the real ones; this follows
\cite{Perspectives}, and $C^\infty(T)$ is the free superfield $C^\infty(N)$
of \cite[Section~3.2]{Derived}. Supertranslations act on $C^\infty(T)$ by
left translation, and the spinor derivatives, generating right
translations, commute with them. Write $\lambda^A$ for the spinor ghosts
and $D_A$ for the paired spinor derivatives, so that $A$ runs over
$\alpha$ in three dimensions, over $\alpha,\dot\alpha$ in four, and over
$i\in\{+,\bpl,-,\bmi\}$ in two. For a dg $\OYd{d}$-module $M$, the associated
multiplet is given by the pure spinor functor of
\cite[Definition~3.5]{Derived},
\begin{equation}
	\A(M)=\bigl(\U\otimes M,\,d_M+\mathcal Q\bigr),\qquad
	\mathcal Q=\textstyle\sum_A\lambda^AD_A.\label{eq:reconstruction}
\end{equation}
We take the spinor ghosts, which are even, to commute with superfields in
$\mathcal Q$; the translation ghosts below are odd, and signs involving them
follow the total parity, so that $v$ anticommutes with odd superfields and
with the spinor derivatives.\footnote{Other
	conventions, such as Koszul signs between ghosts and odd superfields, or
	for the composition of shifted morphisms, change classes such as the
	complex-linear class $c$ of Section~\ref{sec:linear} by an overall sign.
	Since $\Cone(-f)\cong\Cone(f)$, the reconstructed multiplets are
	unaffected.}
Elements of $\U\otimes M$ are $M$-valued superfields
$\Psi=\sum_a f_a(x,\theta)\,m_a$, the pure spinor superfields. The
cohomology of $\A(M)$ in each degree is a space of constrained superfields;
their $\theta$-expansions give the component fields, which are also listed
by the minimal free resolution of $M$ (see below). This is a complex
because $\mathcal Q^2=\tfrac12\sum_{A,B}\lambda^A\lambda^B\{D_A,D_B\}$: the
derivatives act on the superfield factor, while their ghost coefficients
are the torsion quadrics, which vanish in $\OYd{d}$ and hence act by zero
on $M$. In two dimensions, with \eqref{eq:derivatives},
\[
	\mathcal Q=\lp D_++\lbp\bD_++\lm D_-+\lbm\bD_-,\qquad
	\mathcal Q^2=2i\bigl(\lp\lbp\,\partial_{\pplus}+\lm\lbm\,\partial_{\mminus}\bigr).
\]

For a module over the full CE algebra the torsion quadrics need not act by
zero, so $\mathcal Q^2\neq0$ on $\U\otimes M$. Instead, the translation
ghosts act on $M$, and since $M$ is a dg module over $\CE{d}$ this action
satisfies the Leibniz rule $d_Mv+vd_M=dv$: commuting $v$ past the
differential produces multiplication by the torsion quadric $dv$. One
therefore adds to the differential a term $v\partial$, with the
translation ghost paired with the corresponding derivative
\cite[Definition~3.5]{Derived}. Its cross term with $d_M$ in the square of
the differential is $(dv)\partial$, of the same form as $\mathcal Q^2$, and
with the right normalization the two cancel. In two dimensions
\begin{equation}
	\mathscr D_M=d_M+\mathcal Q-2i(\vpp\partial_{\pplus}+\vmm\partial_{\mminus}),
	\label{eq:full-reconstruction}
\end{equation}
and $\mathscr D_M^2=0$ follows from $d_M\vpp+\vpp d_M=\lp\lbp$ and
$d_M\vmm+\vmm d_M=\lm\lbm$, which turn the cross terms into
$-2i(\lp\lbp\,\partial_{\pplus}+\lm\lbm\,\partial_{\mminus})=-\mathcal Q^2$.
The remaining terms vanish: the cross term of $\mathcal Q$ with the
translation-ghost term cancels because $v$ anticommutes with the $D_A$, which
commute with $\partial$, and the square of the translation-ghost term vanishes
because the $v$ anticommute. In four dimensions the added term is
$+2i\,v^{\alpha\dot\alpha}\partial_{\alpha\dot\alpha}$.
We write $\A_{\CE{d}}(M)=(\U\otimes M,\mathscr D_M)$ for this complex.

\paragraph{Gradings.}
Modules are bigraded. The \emph{weight} $w$ is the internal grading in
which $\lambda$ has weight one and $v$ weight two
(Section~\ref{sec:setting}). The \emph{resolution degree} $r$ is the
cohomological degree of a dg module, in which $\lambda$ has degree $0$, $v$
degree $-1$, and the differential $d_M$ degree $+1$. The complex $\A(M)$ is
graded by the \emph{ghost number} $n=r+w$, with superfields in $\U$ of ghost
number zero. Each term of $\mathscr D_M$ raises $n$ by one: $d_M$ through
$r$, $\mathcal Q$ through $w$, and the translation-ghost term through both.
Shifts are written $M[q]^r=M^{r+q}$ in resolution degree and
$M(t)_w=M_{w+t}$ in weight, so that, for instance, the generator of $M(-2)$
has weight two.

As a graded space $\A(M)=\U\otimes_\C M$. Hence $\A$ sends short exact
sequences of modules to short exact sequences of complexes and commutes
with shifts and mapping cones. For quasi-isomorphisms we need a
boundedness hypothesis.

\begin{lemma}\label{lem:qiso}
	$\A$ sends a quasi-isomorphism between modules that are bounded in
	resolution degree to a quasi-isomorphism.
\end{lemma}
\begin{proof}
	Filter by weight. If the resolution degree is bounded, only finitely many
	weights occur in each ghost number, and $\mathcal Q$ and the translation
	terms raise the weight, so the associated graded differential is $d_M$. A
	quasi-isomorphism of modules induces one on $\U\otimes(-)$, and the
	comparison theorem for the spectral sequences of the finite filtrations
	gives the claim \cite[Chapter~5]{Weibel}; compare also
	\cite[Theorem~4.3]{Derived}.
\end{proof}
Every module to which we apply $\A$ is bounded, or is first replaced by a
bounded model. In particular, Ext classes between ordinary modules have
bounded representatives, obtained by truncating a free resolution at a
syzygy module, so derived morphisms reconstruct to morphisms of superfield
complexes \cite{Derived}. We use the standard identification
$\Ext^m_\OY(M,N)=\Hom_{D(\OY)}(M,N[m])$ and do not distinguish a class from
the corresponding derived morphism. A class is fixed only up to the choice of
a representative chain map, and a representative is what enters a cone; the
cone is therefore well defined up to quasi-isomorphism.

In practice this means that short exact sequences and cones of modules
reconstruct to short exact sequences and cones of superfield complexes,
which is how $\A$ is used throughout. A module map reconstructs to a map of
superfields, but not every superfield operator we need arises this way. In
particular, a constraint operator such as $\bD_+\bD_-$ comes in general
only from a derived morphism: there is no module
map from the augmentation module to the chiral module, but there is a
nonzero class in $\Ext^2$, which reconstructs to $\bD_+\bD_-$ up to sign
(Section~\ref{sec:linear}).

\paragraph{Component fields.}
The component fields of $\A(M)$ can be read off from $M$ without solving
any constraint. We state this in two dimensions; in three and four
dimensions $R$ is replaced by the polynomial ring on the $\lambda^A$. At
zero momentum, that is, on superfields independent of
$x$, the spinor derivatives reduce to $\partial/\partial\theta_i$ and
$\mathcal Q$ becomes the Koszul differential
\[
	\mathcal Q_0=\sum_i\lambda_i\frac{\partial}{\partial\theta_i}
\]
on $\C[\theta]\otimes M$, where $\C[\theta]$ is the exterior algebra on
the odd coordinates. Since the ghosts form a regular sequence in $R$, its
cohomology is a Tor group \cite{Perspectives,Derived}:
\[
	H\bigl(\C[\theta]\otimes M,\;d_M+\mathcal Q_0\bigr)=\Tor^R(M,\C).
\]
Concretely, let $F_\bullet\to M$ be a minimal free resolution of an
ordinary graded module $M$ over $R$. A generator of $F_i$ in internal
degree $j$ gives a class in $\Tor_i^R(M,\C)_j$, which in
$\C[\theta]\otimes M$ sits at order $\theta^i$ with module weight $j-i$. It
is therefore a component field at order $i$ in the $\theta$-expansion, of
ghost number $j-i$ before any overall shift, and its parity differs from
that of the lowest component by $i$ modulo two. For example, the chiral
module $M_\Phi=R/(\lp,\lm)$ of Section~\ref{sec:chiral} has the Koszul
resolution $R\leftarrow R(-1)^2\leftarrow R(-2)$, with generators in
internal degrees $0$; $1,1$; and $2$. They give the scalar $\phi$, the two
fermions $\psi_\pm$ and the auxiliary field $F$, all of ghost number zero.
Such counts determine only the graded component spaces; identifying the
multiplet, with its gauge transformations and the relations among its
components, requires the full differential.

\subsection{Cones, constraints and gauge transformations}\label{sec:gauge-cone}
For a chain map $f:M\to N$ of dg CE modules, its mapping cone is
\[
	\Cone(f)=N\oplus M[1],\qquad d(n,m)=(d_Nn+fm,\,-d_Mm),
\]
which sits in the usual long exact sequence in cohomology.

Since $\A$ preserves cones and shifts,
\[
	\A(\Cone(f))\cong\Cone(\A(f)).
\]
Derived module morphisms are handled using bounded representatives as above.

\paragraph{Constraints.}\label{sec:fibers}
Suppose $\A(M)\simeq\mathcal X$ and $\A(N)\simeq\mathcal Y$ are superfields
concentrated in degree zero, and $f$ reconstructs to a differential operator
$L:\mathcal X\to\mathcal Y$. If $L$ is onto, then
\[
	\A(\Cone(f)[-1])\simeq\ker L.
\]
The module-side construction is the shifted cone of $f$, not its ordinary
kernel. Surjectivity of $L$ excludes additional cokernel cohomology, so we
give constraint operators their natural constrained targets rather than all
unconstrained superfields (Section~\ref{sec:linear}).

\paragraph{Gauge transformations.}
If $f$ reconstructs to a superfield gauge transformation $g:\G\to\Pot$,
then $\A(\Cone(f))\simeq[\G\xrightarrow{g}\Pot]_{-1,0}$.
We read this two-term complex as the BRST complex of the gauge symmetry:
degree $0$ contains the potentials, and degree $-1$ the gauge parameters
with the parity shift of the cone, that is, the ghosts. Thus an even scalar
gauge parameter gives an odd ghost. The complex has $H^{-1}=\ker g$, the
parameters that act trivially, and $H^0=\Pot/g\G$, the potentials modulo
gauge.
Conversely, the inverse functor of \cite{Derived} takes derived
supertranslation invariants $C^\bullet(\mathfrak n_d,-)$
\cite[Theorem~4.3]{Derived}. Applied to a superfield gauge transformation,
it gives a morphism of CE modules whose cone reconstructs the gauge complex.
The known transformation thus guides the choice of explicit module
representatives.

For example, the vector multiplet in four dimensions, or the ordinary one
in two, has a real potential $V$ with gauge transformation
$\delta V=i(\bar\Lambda-\Lambda)$, $\Lambda$ chiral. Its kernel consists of
chiral parameters $\Lambda$ that are also real, hence constant. We call
these the \emph{constant gauge transformations}; locally they form a line,
$\C$ in the complexified complex and $\mathbb R$ on the real form. The cone retains them
in degree $-1$, whereas the quotient of potentials by gauge transformations
does not.

In Section~\ref{sec:vectors} each gauge multiplet is obtained this way.
For the two-dimensional vector multiplets we establish, on a connected
local patch, an exact sequence
\begin{equation}
	0\longrightarrow\C\longrightarrow\G\xrightarrow{\ g\ }\Pot
	\xrightarrow{\ \str\ }\{\text{field strengths}\}\longrightarrow0,\label{eq:gauge-exact}
\end{equation}
where $\str$ maps a potential to its field strengths, which are constrained
superfields: potentials modulo gauge are the same as field strengths, and
only the constant gauge transformations act trivially.

\section{Matter multiplets}\label{sec:matter}
Many matter multiplets are defined through constraint superfields; in two-dimensional
$\mathcal N=(2,2)$ supersymmetry these include chiral, twisted-chiral,
semichiral and complex-linear superfields. On the module side, these constraints correspond to modules
of linear support and the maps between them. Such a module
reconstructs the superfields annihilated by the spinor derivatives whose
ghosts survive in it \eqref{eq:linsupp}. Enlarging the support by one ghost
$\lambda_j$ gives a short exact sequence whose connecting
map reconstructs the spinor derivative $D_j$. Enlarging it by several ghosts
one at a time and composing these maps reconstructs products
$D_{j_m}\cdots D_{j_1}$
(Proposition~\ref{prop:yoneda}).

\subsection{Chiral and semichiral multiplets}\label{sec:chiral}\label{sec:semichiral}
From now on we work in two-dimensional $\mathcal N=(2,2)$ supersymmetry. An unconstrained superfield has
$8+8$ complex components and corresponds to the coordinate ring of the
origin of the nilpotence variety:
\[
	k=R/(\lp,\lbp,\lm,\lbm),\qquad \A(k)=\U.
\]

\paragraph{Chiral and twisted-chiral superfields.}
The chiral and twisted-chiral multiplets each have $2+2$ complex off-shell
component fields. Their constraints, and the coordinate rings of the corresponding
planes in the nilpotence variety, are
\[
	\begin{aligned}
		\bD_\pm\Phi       & =0, & M_\Phi & =\C[\lbp,\lbm]=R/(\lp,\lm), \\
		\bD_+\chi=D_-\chi & =0, & M_\chi & =\C[\lbp,\lm]=R/(\lp,\lbm).
	\end{aligned}
\]
For the chiral multiplet the components are a scalar $\phi$, fermions
$\psi_\pm$ and an auxiliary field $F$. Reconstructing the chiral multiplet
is a standard calculation, carried out in detail in
\cite[Section~3.5]{Perspectives}. Partial conjugation,
$\lm\leftrightarrow\lbm$, $D_-\leftrightarrow\bD_-$, interchanges the two
constructions, so the twisted-chiral case follows. The conjugate multiplets work analogously, with
$M_{\bar\Phi}=\C[\lp,\lm]$ and $M_{\bar\chi}=\C[\lp,\lbm]$. We write $\Ch$
and $\bar\Ch$ for the sheaves of chiral and antichiral superfields.

\paragraph{Semichiral superfields.}
A left semichiral superfield obeys only one spinor constraint. Its module
is the coordinate ring of an axis rather than a plane:
\[
	\bD_+\mathbb X_L=0,\qquad
	M_{\mathbb X_L}=\C[\lbp]=R/(\lp,\lm,\lbm).
\]
Here reconstruction can be computed directly:
\[
	\A(M_{\mathbb X_L})
	=\bigl(\U\otimes_\C\C[\lbp],\;\mathcal Q=\lbp\bD_+\bigr).
\]
Explicitly, $\mathcal Q(\lbp^{\,n}U)=\lbp^{\,n+1}\bD_+U$, so a cocycle of degree
$n$ is $\lbp^{\,n}U$ with $\bD_+U=0$. In degree zero there are no
coboundaries, so $H^0=\ker\bD_+$. In positive degree every cocycle is exact,
since $\bD_+(\bar\theta^+U)=U$ whenever $\bD_+U=0$:
\[
	\mathcal Q\bigl(\lbp^{\,n-1}\bar\theta^+U\bigr)=\lbp^{\,n}U,
	\qquad n\ge1.
\]
Thus reconstruction gives the left semichiral multiplet, with $4+4$ complex
components, in degree zero and no other cohomology. The right-handed and
conjugate constructions are analogous, with $M_{\mathbb X_R}=\C[\lbm]$,
$M_{\bar{\mathbb X}_L}=\C[\lp]$ and $M_{\bar{\mathbb X}_R}=\C[\lm]$.

\subsection{Linear support and derived constraints}\label{sec:yoneda}

The examples of Section~\ref{sec:chiral} suggest a geometric
interpretation of first-order superfield constraints: the modules
are structure sheaves of linear subspaces of the nilpotence variety,
and the corresponding superfields are annihilated by the spinor
derivatives along those subspaces.

First-order constraints do not exhaust the possibilities. For
example, the complex-linear constraint
$\bD_+\bD_-\Sigma=0$ of Section~\ref{sec:linear} requires the
product of two derivatives to vanish, without imposing either
constraint separately. Such higher-order operators will arise
from derived morphisms between the structure sheaves of linear
subspaces, and their compositions from Yoneda products.

\paragraph{Linear subspaces and first-order constraints.}
We now make this interpretation precise: the structure sheaf of a
linear subspace of the nilpotence variety reconstructs to the superfields
annihilated by the spinor derivatives along that subspace
(Proposition~\ref{prop:linsupp}). The results of this subsection hold in
general, regardless of the dimension and the amount of supersymmetry. Here
$Y$ denotes the nilpotence variety and $\OY$ its ring of functions; however,
most of the examples we discuss are two-dimensional.
Let $\nodd$ denote the odd part of the complexified supertranslation
algebra, with bracket
\[
    [-,-]\colon \operatorname{Sym}^2\nodd\longrightarrow\mathfrak t,
\]
where $\mathfrak t$ is the space of translations. The nilpotence
variety $Y\subset\nodd$ is defined by the equation $[u,u]=0$.

Consider a vector subspace $\mathfrak l\subset\nodd$ contained in $Y$.
By polarization, this is equivalent to
\[
    [u,v]=0\qquad\text{for all }u,v\in \mathfrak l.
\]
Thus $\mathfrak l$ is an abelian subspace of odd supertranslations.
Its inclusion defines a closed linear subscheme
$Z_{\mathfrak l}\hookrightarrow Y$, with structure sheaf
\[
    \calO_{Z_{\mathfrak l}}=\operatorname{Sym}(\mathfrak l^*),
\]
regarded as an $\OY$-module through restriction of polynomial
functions from $Y$ to $Z_{\mathfrak l}$.

For $u\in\nodd$, write $D_u$ for the corresponding spinor
derivative on superspace. The derivatives associated with $\mathfrak l$
mutually anticommute, and therefore define an integrable system
of first-order constraints. We denote its solution space by
\[
    \ker D_{\mathfrak l}
    :=\{\Phi\in\U:\ D_u\Phi=0\text{ for every }u\in \mathfrak l\}.
\]

\begin{proposition}\label{prop:linsupp}
    Let $\mathfrak l\subset Y$ be a linear subspace of the nilpotence
    variety. Then the reconstruction of its structure sheaf
    is quasi-isomorphic to the superfield satisfying the
    corresponding first-order constraints:
    \begin{equation}
        H^\bullet\bigl(\A(\calO_{Z_{\mathfrak l}})\bigr)
        =
        \begin{cases}
            \ker D_{\mathfrak l}, & \bullet=0,\\
            0, & \bullet\ne0.
        \end{cases}
        \label{eq:linsupp}
    \end{equation}
    The natural inclusion
    $\ker D_{\mathfrak l}\hookrightarrow\A(\calO_{Z_{\mathfrak l}})$
    is a supersymmetric quasi-isomorphism.

    Conversely, if a bounded module $M$ reconstructs to
    $\ker D_{\mathfrak l}$, viewed as a complex concentrated in degree zero,
    through a quasi-isomorphism of multiplets in the sense of
    \cite{Derived}, that is, in the category on which the inverse functor of
    \cite[Theorem~4.3]{Derived} is defined, then
    $M\simeq\calO_{Z_{\mathfrak l}}$ in the derived category.
\end{proposition}

\begin{proof}
    Choose a basis $u_1,\ldots,u_r$ of $\mathfrak l$ and let
    $t^1,\ldots,t^r$ be the dual linear coordinates.
    Since $\calO_{Z_{\mathfrak l}}=\C[t^1,\ldots,t^r]$, the
    reconstruction differential becomes
    \[
        \mathcal Q_{\mathfrak l}=\sum_{a=1}^r t^aD_{u_a}.
    \]
    The condition $\mathfrak l\subset Y$ implies
    $\{D_{u_a},D_{u_b}\}=0$ for all $a,b$.

    By the super Frobenius theorem, these commuting odd
    derivatives can be simultaneously straightened.
    In adapted local superspace coordinates
    $(\eta^1,\ldots,\eta^r,y)$, with the $\eta^a$ odd,
    they take the form
    \[
        D_{u_a}=\frac{\partial}{\partial\eta^a},
        \qquad
        \mathcal Q_{\mathfrak l}
        =\sum_{a=1}^r t^a
          \frac{\partial}{\partial\eta^a}.
    \]
    In flat superspace these adapted coordinates may
    be chosen globally.

    Introduce the contracting operator
    \[
        h=\sum_{a=1}^r
          \eta^a\frac{\partial}{\partial t^a}.
    \]
    Its graded commutator with $\mathcal Q_{\mathfrak l}$ is
    \[
        \mathcal Q_{\mathfrak l}h+h\mathcal Q_{\mathfrak l}
        =
        \sum_{a=1}^r
        \left(
            t^a\frac{\partial}{\partial t^a}
            +\eta^a\frac{\partial}{\partial\eta^a}
        \right).
    \]
    The right-hand side counts the total polynomial
    degree in the variables $t^a$ and $\eta^a$.
    It is invertible in every strictly positive degree,
    so the complex contracts onto the functions
    independent of all $t^a$ and $\eta^a$.

    These are precisely the superfields satisfying
    $D_{u_a}\Phi=0$ for every $a$. They lie in ghost
    number zero, proving \eqref{eq:linsupp}.
    Their inclusion is supersymmetric because the
    operators $D_{u_a}$ commute with left
    supertranslations.

    Finally, the inverse equivalence of
    \cite[Theorem~4.3]{Derived}, applied to the
    equivariant quasi-isomorphism
    $\A(M)\simeq\A(\calO_{Z_{\mathfrak l}})$, gives
    $M\simeq\calO_{Z_{\mathfrak l}}$ in the derived category.
\end{proof}

In two dimensions, let $u_i\in\nodd$, $i\in\{+,\bpl,-,\bmi\}$, be the basis
dual to the ghosts $\lambda_i$, so that $D_{u_i}=D_i$. The span of some of the
$u_i$ lies in $Y$ exactly when it contains no torsion pair, and its ring is
then the polynomial ring on the corresponding ghosts. The zero subspace, an
axis and a plane give the unconstrained, semichiral, and chiral or
twisted-chiral modules of Section~\ref{sec:chiral}.

\begin{remark}\label{rem:linsupp}
	The scheme structure matters: Proposition~\ref{prop:linsupp} needs the
	reduced ring $\calO_{Z_{\mathfrak l}}$, not merely a module supported on
	$Z_{\mathfrak l}$. For example, $M=\C[\lp,\lm]/(\lm^2)$, with $\lbp,\lbm$
	acting by zero, is supported on the $\lp$-axis. Since
	$\lm M\cong\C[\lp](-1)$, there is a short exact sequence
	\[
		0\to\C[\lp](-1)\xrightarrow{\ \lm\ }M\to\C[\lp]\to0,
	\]
	and, as in the proof of Proposition~\ref{prop:yoneda}(i), its connecting
	map is $D_-$ acting on $\ker D_+$. This map is neither injective nor onto,
	so $\A(M)$ has cohomology $\ker D_+\cap\ker D_-$ in degree zero and the
	cokernel of $D_-$ in degree one. The thickening in the $\lm$ direction
	imposes the additional constraint $D_-\Phi=0$ and adds cohomology in
	degree one, instead of reconstructing the semichiral constraint
	$D_+\Phi=0$.
\end{remark}

\paragraph{Yoneda products.}
Enlarging the subspace imposes additional constraints. For linear subspaces
$\mathfrak l\subset\mathfrak l'\subset Y$, restriction
$\calO_{Z_{\mathfrak l'}}\to\calO_{Z_{\mathfrak l}}$ reconstructs the inclusion
$\ker D_{\mathfrak l'}\subset\ker D_{\mathfrak l}$. If
$\mathfrak l'=\mathfrak l\oplus\C u$, let $\ell\in\mathfrak l'^*$ be the linear
form with $\ell(\mathfrak l)=0$ and $\ell(u)=1$. Then
\begin{equation}
	0\to\calO_{Z_{\mathfrak l'}}(-1)\xrightarrow{\ \ell\ }\calO_{Z_{\mathfrak l'}}
	\to\calO_{Z_{\mathfrak l}}\to0.\label{eq:yoneda-step}
\end{equation}
Its connecting class $e_u$, that is, the derived morphism
$\calO_{Z_{\mathfrak l}}\to\calO_{Z_{\mathfrak l'}}(-1)[1]$ in
$\Ext^1(\calO_{Z_{\mathfrak l}},\calO_{Z_{\mathfrak l'}}(-1))$, reconstructs
the derivative $D_u$; in two dimensions we write $e_i=e_{u_i}$. For a flag
$\mathfrak l=\mathfrak l_0\subset\mathfrak l_1\subset\cdots\subset\mathfrak l_m=\mathfrak l'$
with $\mathfrak l_k=\mathfrak l_{k-1}\oplus\C u_k$, write
$\calO_k=\calO_{Z_{\mathfrak l_k}}$. The Yoneda product of the classes
$e_{u_k}$, obtained by splicing the short exact sequences or, equivalently,
by composing them as morphisms
$\calO_0\to\calO_1(-1)[1]\to\cdots\to\calO_m(-m)[m]$ in the derived category
\cite[\S3.4]{Weibel}, reconstructs the product of the derivatives, rather
than their simultaneous constraints.

\begin{proposition}\label{prop:yoneda}
	Let $\mathfrak l\subset\mathfrak l'\subset Y$ be linear subspaces.
	\begin{enumerate}[label=(\roman*)]
		\item If $\mathfrak l'=\mathfrak l\oplus\C u$, the class $e_u$ reconstructs to
		      $D_u\colon\ker D_{\mathfrak l}\to\Pi\ker D_{\mathfrak l'}$, where $\Pi$
		      reverses parity. This map is onto with kernel $\ker D_{\mathfrak l'}$.
		\item For a flag from $\mathfrak l$ to $\mathfrak l'$ as above, the Yoneda product
		      $e=e_{u_m}\circ\cdots\circ e_{u_1}\colon\calO_0\to\calO_m(-m)[m]$
		      reconstructs to
		      $\pm D_{u_m}\cdots D_{u_1}\colon\ker D_{\mathfrak l}\to\Pi^m\ker D_{\mathfrak l'}$.
		      This map is onto, so $e\neq0$ and
		      \[
			      \A\bigl(\Cone(e)\bigr)\simeq\{X\in\ker D_{\mathfrak l}:\ D_{u_m}\cdots D_{u_1}X=0\}[1].
		      \]
	\end{enumerate}
\end{proposition}
\begin{proof}
	(i) Lift $U\in\ker D_{\mathfrak l}$ to the middle term of the reconstructed
	sequence \eqref{eq:yoneda-step}. In a basis of $\mathfrak l'$ extending one
	of $\mathfrak l$ by $u$, applying $\mathcal Q$ gives $\ell\,D_uU$, identifying
	the connecting map with $D_u$. Since $\mathfrak l'\subset Y$, the derivatives
	$D_v$, $v\in\mathfrak l'$, anticommute and square to zero, so $D_uU$ lies in
	$\ker D_{\mathfrak l'}$. In the adapted coordinates of the proof of
	Proposition~\ref{prop:linsupp}, applied to $\mathfrak l'$, $D_u$ is the
	derivative along an odd coordinate $\eta$; multiplying a target field by
	$\eta$ gives a preimage in $\ker D_{\mathfrak l}$, proving surjectivity. The
	kernel is $\ker D_{\mathfrak l'}$.
	(ii) The product is the derived morphism
	$\calO_0\to\calO_m(-m)[m]$. The internal shift $(-m)$ raises the weight
	of the target generator by $m$, while $[m]$ lowers its resolution degree
	by $m$. They therefore cancel in ghost number $n=r+w$: the reconstructed
	source and target both have cohomology in degree zero. The parity reversal
	$\Pi^m$ is separate from this cancellation and records the parity of the
	$m$ spinor derivatives.
	The Yoneda product is composition in the derived category. The classes
	are represented by roofs of bounded modules whose backward arrows are
	quasi-isomorphisms; by Lemma~\ref{lem:qiso}, $\A$ sends these to roofs
	of superfield complexes with quasi-isomorphic backward arrows, so it is
	well defined on derived morphisms and compatible with composition. The
	reconstructed composite is the product of the operators in (i), up to the
	signs from the parity shifts. Multiplying a target field by the $m$ odd
	coordinates along $u_1,\dots,u_m$ gives a preimage, so the product is onto. A morphism
	between complexes concentrated in one degree is zero in the derived category
	only if it is zero as a map of superfields, so $e\neq0$. Since the map is onto,
	the cone is the shifted kernel (Section~\ref{sec:fibers}).
\end{proof}

On $\ker D_{\mathfrak l}$ one has $D_{u+v}=D_u$ for $v\in\mathfrak l$, and the
$D_{u_k}$ anticommute, so the operator in (ii) depends on the flag only
through $u_1\wedge\cdots\wedge u_m\in\det(\mathfrak l'/\mathfrak l)$.
Taking $\mathfrak l=0$ and $\mathfrak l'=\C u_i$ in two dimensions and applying
the cone construction of Section~\ref{sec:gauge-cone}, we obtain a cone
description of the semichiral constraint, also used in Section~\ref{sec:lvm}:
\begin{equation}
	\A(\C[\lambda_i])[1]\simeq\Cone\bigl(D_i:\U\to\Pi\ker D_i\bigr)\simeq(\ker D_i)[1].
	\label{eq:axis}
\end{equation}

The degree of the class is $m=\dim\mathfrak l'/\mathfrak l$, at most the
dimension of a linear subspace of $Y$. This is two for $\mathcal N=(2,2)$ and
for $4d$ $\mathcal N=1$, so second order is the maximum there. With more
supercharges higher orders occur: for $4d$ $\mathcal N=2$ the subspace on
which all $\lambda^\alpha_i$ vanish is linear and four-dimensional, and $m=4$
gives the constraint $\bar D^4X=0$.

\paragraph{General $m$.}
For general $m$, in the situation of Proposition~\ref{prop:yoneda}(ii),
introduce auxiliary fields $S_k\in\Pi^k\ker D_{\mathfrak l_k}$, $1\le k<m$.
Composing the classes $e_{u_k}$ amounts to taking the cone over the staircase
\[
	\begin{tikzcd}[column sep=large]
		\calO_0 \arrow[r,"e_{u_1}"] & \calO_1(-1)[1] \\
		& \calO_1(-1)[1] \arrow[u,"1"] \arrow[r,"e_{u_2}"] & \calO_2(-2)[2]
	\end{tikzcd}
\]
shown for $m=2$. The vertical arrow is the identity of the middle module; it
is what makes the total complex compute the composite rather than the two
classes separately. The differential is
$\begin{pmatrix}e_{u_1}&-1\\0&e_{u_2}\end{pmatrix}$, which for $m=2$ is
\eqref{eq:module-L}. Applying $\A$, the constraint
$D_{u_m}\cdots D_{u_1}X=0$ is equivalent to the first-order system
\[
	\begin{pmatrix}
		D_{u_1} & -1      &        &         \\
		        & D_{u_2} & \ddots &         \\
		        &         & \ddots & -1      \\
		        &         &        & D_{u_m}
	\end{pmatrix}
	\begin{pmatrix}X\\ S_1\\ \vdots\\ S_{m-1}\end{pmatrix}=0,
\]
that is, $S_1=D_{u_1}X$, $S_{k+1}=D_{u_{k+1}}S_k$ and $D_{u_m}S_{m-1}=0$;
each entry is well defined by Proposition~\ref{prop:yoneda}(i).
Eliminating $S_1,\dots,S_{m-1}$ by row and column operations as in
\eqref{eq:elimination} below splits off $m-1$ identities and leaves the Schur
complement $D_{u_m}\cdots D_{u_1}$. On the module side the diagonal entries
are the classes $e_{u_k}$ and the entries above it are identities of the
intermediate modules $\calO_k(-k)[k]$; the elimination leaves the
Yoneda product $e$ of Proposition~\ref{prop:yoneda}(ii).

\subsection{Complex-linear superfields}\label{sec:linear}
A complex-linear superfield satisfies $\bD_+\bD_-\Sigma=0$. The operator
$\bD_+\bD_-\colon\U\to\Ch$ is onto, so the constraint removes $2+2$ of the
$8+8$ unconstrained components, leaving $6+6$. A second-order constraint
is not imposed by a module map. Proposition~\ref{prop:yoneda} immediately
gives the relevant complex instead: the two first-order classes $e_{\bpl}$
and $e_{\bmi}$ compose to the degree-two class $c$, and complex-linear
matter is its cone, shifted by one, as in \eqref{eq:MSigma}. Applying $\A$
to the two-step complex $L$ of that equation reconstructs the
first-order formulation \eqref{eq:first-order}: the intermediate module
$\C[\lbp](-1)$ becomes an auxiliary semichiral field, and eliminating that
field returns $\bD_+\bD_-$. We record the complex first, then its
reconstruction.

\paragraph{The module complex.}
Every ingredient of this formulation comes from a module. The three kinds of
superfields are reconstructed from $k$, $\C[\lbp]$ and $M_\Phi$, and each
derivative from dropping one ghost (Proposition~\ref{prop:yoneda} with
the flag $0\subset\C u_{\bpl}\subset\C u_{\bpl}\oplus\C u_{\bmi}$), that is, from the
sequences \eqref{eq:yoneda-step}
\[
	0\to\C[\lbp](-1)\xrightarrow{\ \lbp\ }\C[\lbp]\xrightarrow{\ \epsilon\ }k\to0,
	\qquad
	0\to M_\Phi(-1)\xrightarrow{\ \lbm\ }M_\Phi\xrightarrow{\ r\ }\C[\lbp]\to0,
\]
where $\lbp,\lbm$ denote multiplication, $\epsilon$ is the augmentation and
$r$ the restriction $f(\lbp,\lbm)\mapsto f(\lbp,0)$. Their classes
\[
	e_{\bpl}\colon k\to\C[\lbp](-1)[1],\qquad
	e_{\bmi}\colon\C[\lbp](-1)[1]\to M_\Phi(-2)[2],
\]
the second twisted by $(-1)[1]$, reconstruct to $\bD_+$ and $\bD_-$
(Proposition~\ref{prop:yoneda}(i)). The first-order operator \eqref{eq:first-order} below
is the reconstruction of
\begin{equation}
	L=\begin{pmatrix}e_{\bpl} & -1 \\ 0 & e_{\bmi}\end{pmatrix}\colon\
	k\oplus \C[\lbp](-1)[1] \longrightarrow \C[\lbp](-1)[1] \oplus M_\Phi(-2)[2].
	\label{eq:module-L}
\end{equation}
Row and column operations of this kind, as in \eqref{eq:elimination} below,
make sense in the derived category as well, with $e_{\bpl}$ and $e_{\bmi}$
in place of the derivatives. They split off the identity of the middle
module $E=\C[\lbp](-1)[1]$, whose cone
vanishes, and leave the Yoneda product. This defines the complex-linear
module:
\begin{equation}
	c:=e_{\bmi}\circ e_{\bpl}\colon\
	k\to\C[\lbp](-1)[1]\to M_\Phi(-2)[2],\qquad
	M_\Sigma[1]:=\Cone(c)\simeq\Cone(L).
	\label{eq:MSigma}
\end{equation}

Concretely, the cone in \eqref{eq:MSigma} is the splice of the two sequences. Twisted by
$(-1)$, the second ends where the first begins; the common module
$\C[\lbp](-1)$, the one carrying the auxiliary field, is dropped and the maps through it are composed,
\[
	\begin{tikzcd}[column sep=small]
		0\arrow[r] & M_\Phi(-2)\arrow[r,"\lbm"] & M_\Phi(-1)\arrow[rr,"\lbp\circ r"]\arrow[dr,two heads,"r"']
		& & \C[\lbp]\arrow[r,"\epsilon"] & k\arrow[r] & 0.\\
		& & & \C[\lbp](-1)\arrow[ur,hook,"\lbp"'] & & &
	\end{tikzcd}
\]
This gives the exact sequence
\begin{equation}
	0\to M_\Phi(-2)\xrightarrow{\ \lbm\ }M_\Phi(-1)
	\xrightarrow{\ \lbp\circ r\ }\C[\lbp]\xrightarrow{\ \epsilon\ }k\to0,
	\label{eq:linear-splice}
\end{equation}
exact at the splice because $\ker(\lbp\circ r)=\ker r=\operatorname{im}\lbm$
and $\operatorname{im}(\lbp\circ r)=\operatorname{im}\lbp=\ker\epsilon$. Its
class is $c$. On the resolution of $k$, $c$ is, up to sign, the cocycle
dual to the syzygy $K_{\bpl\bmi}$ (Appendix~\ref{app:linear-class}).

Over $R$ the cone has an explicit representative: $c$ is realized by the
quotient $K_R\to K_R/L_R\simeq M_\Phi(-2)[2]$ of the Koszul complex, so
$\Cone(c)\simeq L_R[1]$ is the complex with free ranks $(1,4,5,2)$ and
linear differentials. Its twelve generators are the $6+6$ component fields
of complex-linear matter (Appendix~\ref{app:finite-model}).

\paragraph{A first-order system.}
Applying $\A$ to $L$ gives a useful presentation of the complex-linear
multiplet: the modules $k$,
$\C[\lbp](-1)[1]$ and $M_\Phi(-2)[2]$ become $\U$, $\Pi\ker\bD_+$ and
$\Ch$. The intermediate module is thus the module
of an auxiliary field, and the reconstruction is the first-order operator
\begin{equation}
	\begin{gathered}
		\mathcal L:=\A(L)=\begin{pmatrix}\bD_+ & -1\\ 0 & \bD_-\end{pmatrix}\colon\
		\U\oplus\Pi\ker\bD_+\longrightarrow\Pi\ker\bD_+\oplus\Ch,\\
		\mathcal L\begin{pmatrix}\Sigma\\ S\end{pmatrix}
		=\begin{pmatrix}\bD_+\Sigma-S\\ \bD_-S\end{pmatrix}.
	\end{gathered}
	\label{eq:first-order}
\end{equation}
Each entry is first order, and each is well defined: $\bD_+$ lands in
semichiral superfields since $\bD_+^2=0$, and $\bD_-$ maps them to chiral
ones since $\bD_+\bD_-=-\bD_-\bD_+$. The first row says $S=\bD_+\Sigma$, the
second $\bD_-S=0$. Because the entry $-1$ is invertible, $S$ can be
eliminated by row and column operations:
\begin{equation}
	\begin{pmatrix}1&0\\ \bD_-&1\end{pmatrix}
	\mathcal L
	\begin{pmatrix}1&0\\ \bD_+&1\end{pmatrix}
	=\begin{pmatrix}0&-1\\ \bD_-\bD_+&0\end{pmatrix}.
	\label{eq:elimination}
\end{equation}
The outer matrices are invertible, so $\mathcal L$ is equivalent to the identity on
the auxiliary pair, which carries no information, plus the Schur complement
\[
	0-\bD_-(-1)^{-1}\bD_+=\bD_-\bD_+=-\bD_+\bD_-\colon\U\to\Ch.
\]
In particular $\ker\mathcal L\cong\ker(\bD_+\bD_-)$, via
$\Sigma\mapsto(\Sigma,\bD_+\Sigma)$. The auxiliary field also organizes the
multiplet: $S=\bD_+\Sigma$ is chiral, its kernel consists of the semichiral
superfields, and $\bar\theta^+\Phi\mapsto\Phi$ for chiral $\Phi$, so
complex-linear matter is an extension of chiral by semichiral matter,
$6+6=(4+4)+(2+2)$.

Finally, reconstructing the splice carries out the elimination
\eqref{eq:elimination} on superfields. Let $p_1$ and $p_0$ be the generators of $M_\Phi(-1)$ and $M_\Phi(-2)$, so
that the map $\lbm$ sends $p_0\mapsto\lbm p_1$. The reconstruction of $c$ is the
connecting map of the reconstructed sequence \eqref{eq:linear-splice}:
lift, apply $\mathcal Q$, pull back, once for each step. Lifting
$\Sigma\in\U$ to $\Sigma\cdot1\in\U\otimes\C[\lbp]$, where
$\mathcal Q=\lbp\bD_+$, produces $S$:
\[
	\mathcal Q(\Sigma\cdot1)=\lbp\,\bD_+\Sigma=(\lbp\circ r)(S\,p_1).
\]
Lifting $S\,p_1$ to $\U\otimes M_\Phi(-1)$, where
$\mathcal Q=\lbp\bD_++\lbm\bD_-$, and using $\bD_+S=0$, produces $\bD_-S$:
\[
	\mathcal Q(S\,p_1)=\lbm\bigl((\bD_-S)\,p_0\bigr).
\]
The two steps are the connecting maps of $e_{\bpl}$ and $e_{\bmi}$
(Proposition~\ref{prop:yoneda}(i)), and together they give the Schur
complement of \eqref{eq:elimination}:
\[
	\A(c)\colon\ \Sigma\longmapsto\bD_-\bD_+\Sigma=-\bD_+\bD_-\Sigma,
	\qquad\U\to\Ch.
\]
The shifts $(-2)[2]$ put $p_0$ in ghost number zero, with character
opposite to that of $\bD_+\bD_-$, so this map is equivariant. It is onto,
hence
\[
	\A(M_\Sigma)\simeq\ker\bigl(\bD_+\bD_-\colon\U\to\Ch\bigr).
\]
Imposing $\bD_+$ and $\bD_-$ separately would instead use the quotient map
$M_\Phi\to k$, which reconstructs to $\Ch\subset\U$; the cone imposes only
the weaker product constraint.

\paragraph{Remarks.}
Partial conjugation $\lm\leftrightarrow\lbm$, $D_-\leftrightarrow\bD_-$
gives the twisted complex-linear multiplet $\bD_+D_-\Xi=0$, from the
analogous class $k\to M_\chi(-2)[2]$; conjugation gives
$D_+D_-\bar\Sigma=0$. The same construction in four dimensions reconstructs
$\bar D^2\Sigma=0$, up to normalization. In two dimensions $c$ is also realized by a finite strict
$\CE{2}$-module, with the translation ghosts acting as explicit homotopies
(Appendix~\ref{app:finite-model}).

\section{Matter constraints on curvature}\label{sec:gauge}
Coupling the matter multiplets of Section~\ref{sec:matter} to a gauge field
while preserving supersymmetry is constrained. The matter superfields obey
constraints such as $\bD_\pm\Phi=0$, and their covariant versions, such as
$\bar\nabla_\pm\Phi=0$, are consistent only if the curvature of the gauge
connection satisfies matching conditions. On the module side, these
restrictions tell us which quadratic ghost expressions to set to zero,
giving the rings of the gauge multiplets considered below.

\subsection{Curvature inside the CE algebra}\label{sec:curv-ce}
We first consider quotients $M=R/J$ by quadratic monomials, with $J$ containing
$I_T$. In the shifted reconstruction $\A(M)[1]$, weight zero gives a gauge
parameter $\kappa$ and weight one a spinor connection
$\Psi=\sum_i\lambda_i\Gamma_i$, with $\delta\Gamma_i=D_i\kappa$.
Weight two records its curvature constraints, and the ring determines which
curvature components vanish.

\paragraph{Connection and curvature.}
For an abelian connection, set $\nabla_i=D_i+i\Gamma_i$ and
$\nabla_r=\partial_r+i\Gamma_r$, where the $\Gamma_i$ are odd superfields.
The covariant derivatives satisfy
\[
	\{\nabla_i,\nabla_j\}=T^r_{ij}\nabla_r+iF_{ij}.
\]
Expanding this bracket gives the spinor curvature
\[
	\mathscr F=\tfrac12\sum_{i,j}\lambda_i\lambda_jF_{ij},\qquad
	F_{ij}=D_i\Gamma_j+D_j\Gamma_i-T^r_{ij}\Gamma_r,
\]
where $\Gamma_r$, $r\in\{\pplus,\mminus\}$, is the vector connection and the
only nonzero torsion is $T^{\pplus}_{+\bpl}=T^{\mminus}_{-\bmi}=2i$. The
components with a vector index, of weights three and four, are fixed by the
spinor ones through the conventional constraints and the Bianchi identities,
so only $\mathscr F$ matters below. The reconstruction differential produces
$\mathscr F$ without its torsion term:
\begin{equation}
	\mathcal Q\Psi=\tfrac12\sum_{i,j}\lambda_i\lambda_j\,(D_i\Gamma_j+D_j\Gamma_i).
	\label{eq:QPsi}
\end{equation}
For $\lambda_i,\lambda_j$ not a torsion pair, $F_{ij}=D_i\Gamma_j+D_j\Gamma_i$;
in particular $F_{ii}=2D_i\Gamma_i$.

\paragraph{Torsion pairs: conventional constraints.}
For the two torsion pairs,
\[
	F_{+\bpl}=D_+\Gamma_{\bpl}+\bD_+\Gamma_+-2i\Gamma_{\pplus},\qquad
	F_{-\bmi}=D_-\Gamma_{\bmi}+\bD_-\Gamma_--2i\Gamma_{\mminus}.
\]
Setting these components to zero defines the vector connection
$\Gamma_{\pmpm}$ in terms of the spinor connection, rather than imposing
additional equations on it. The exact CE quadrics record these torsion
terms: since $\lp\lbp=d\vpp$ is exact, the
$\lp\lbp$ term of the curvature combines with the translation-ghost
component of the connection into $F_{+\bpl}$, and likewise for $\lm\lbm$.
Passing to $\OY$ removes these torsion-pair terms after the conventional
constraints have defined the vector connection; they impose no additional
conditions on the spinor connection.

\paragraph{The other quadrics: a quotient ring is a constraint system.}
In its lowest degrees, $\A(M)[1]$ is
\[
	\underbrace{\kappa}_{\text{degree }-1}\;\longmapsto\;
	\underbrace{\Psi=\sum_i\lambda_i\Gamma_i}_{\text{degree }0}\;\longmapsto\;
	\underbrace{\mathcal Q\Psi}_{\text{degree }1},
\]
with $\kappa$ an even superfield. The first map gives
$\mathcal Q\kappa=\sum_i\lambda_iD_i\kappa$, reproducing the gauge
transformation $\delta\Gamma_i=D_i\kappa$. The second gives the
spinor-derivative terms in the curvature, as in \eqref{eq:QPsi}.
Since $J$ has no linear part, $M_1=R_1$ and
all four spinor-connection components are present. In degree one $\mathcal Q\Psi$ lies in
$\U\otimes M_2$, so by \eqref{eq:QPsi} the cocycle condition imposes
$F_{ij}=0$ exactly for the quadrics $\lambda_i\lambda_j$ that survive in $M$.
Thus degree-zero cocycles are connections satisfying these curvature
constraints, and boundaries are gauge variations.

For example, if $\lbp^2$ survives in $M$, its coefficient in
$\mathcal Q\Psi$ must vanish, imposing $\bD_+\Gamma_{\bpl}=0$.
By contrast, if $\lbp\lm\in J$, its class is zero in $M$, so the cocycle
equation does not impose $F_{\bpl-}=0$.
Each quadric therefore plays one of three roles:
\begin{center}\small
	\begin{tabular}{ll}\toprule
		quadric $\lambda_i\lambda_j$ & role                                                 \\\midrule
		torsion pair, in $I_T$       & conventional constraint, solved for $\Gamma_{\pmpm}$ \\
		survives in $M$              & constraint $F_{ij}=0$                                \\
		in $J$ but not in $I_T$      & allowed curvature component $F_{ij}$                 \\\bottomrule
	\end{tabular}
\end{center}
Adding a quadric $\lambda_i\lambda_j$ to $J$ sets its class to zero in $M$,
removing the equation $F_{ij}=0$; we then call $F_{ij}$ an allowed curvature
component, and identify it with the quadric $\lambda_i\lambda_j$.
The higher degrees of $\A(M)[1]$ carry the Bianchi identities, and
Proposition~\ref{prop:strengths} shows that for the rings below they add no
further local cohomology.

A linear element of $J$ has a different effect. If $J$ is generated by linear
expressions, as for the matter modules of Section~\ref{sec:matter}, then
$\A(R/J)$ is concentrated in degree zero by \eqref{eq:linsupp}, and
$\A(R/J)[1]$ has no connection at all. Retaining all spinor-connection
components in this construction requires $(R/J)_1=R_1$; their curvature
constraints are encoded in the quadratic part $J_2$. This is why
quadratically generated ideals are the natural class below.

\paragraph{Three and four dimensions.}
In $\CE{3}$ every quadric $\lambda^\alpha\lambda^\beta=dv^{\alpha\beta}$ is
exact, and in $\CE{4}$ the mixed quadrics $\lambda^\alpha\bar\lambda^{\dot\alpha}$
are, while $\lambda^\alpha\lambda^\beta$ and
$\bar\lambda^{\dot\alpha}\bar\lambda^{\dot\beta}$ survive in $\OYd{4}$ and give
$F_{\alpha\beta}=F_{\dot\alpha\dot\beta}=0$. The field strengths sit in the
weight-three classes of Section~\ref{sec:setting}: the curvature of the $3d$
vector multiplet is, up to normalization, $\zeta^\alpha W_\alpha$ plus
descendants \cite{Superspace}, and in four dimensions the two families carry
$W_\alpha$ and $\bar W_{\dot\alpha}$. These classes vanish in $\OYd{d}$, but the
module still records them: the free resolution
$0\to R(-3)^2\to R(-2)^3\to R\to\OYd{3}$ has two cubic syzygies, which
correspond to $\zeta^\alpha$ and give the gaugino. In two dimensions
$\CE{2}\simeq\OY$ has no such classes. The spinor-curvature components are
quadratic, while the vector-index components are determined by the
conventional constraints and Bianchi identities. Thus the quotient rings
considered here encode their spinor-curvature constraints.

\subsection{The selection principle}\label{sec:selection}
We start with chiral matter. To gauge a symmetry acting on chiral superfields
one needs covariant derivatives for which $\bar\nabla_\pm\Phi=0$ is
consistent, that is $\{\bar\nabla_+,\bar\nabla_-\}=\bar\nabla_\pm^2=0$. This
is the familiar statement that some spinor curvatures must vanish for
covariantly chiral superfields to exist; in pure spinor language it reads
$\bigl(\sum_i\lambda_i\nabla_i\bigr)^2=0$ on $\U\otimes M_\Phi$. We turn it around:
given the matter, which curvature is allowed? Geometric frameworks in which
the odd distribution of superspace and the filtration it generates are the
basic data are developed in \cite{HahnerSaberi}; the selection below is an
elementary, abelian and flat calculation in this spirit.

A real connection acting on charged matter acts on its conjugate with the
opposite charge. Since we work with complexified fields, we impose this
explicitly: a matter collection always contains, together with each module,
its conjugate. This hypothesis is
essential, as we explain after the table below.

Let one abelian connection act on several charged matter modules. For these
integrability conditions, the value of a nonzero charge is irrelevant;
we take charges $\pm1$ for notational simplicity:
$\nabla_A=D_A\pm i\Gamma_A$, with opposite signs on a module and its
conjugate, and the connection term acts by multiplication. Conventional elimination of the vector connection gives
$\mathcal Q\Psi=\mathscr F$ modulo $I_T$. For a homogeneous ideal $H$ let
$H_2$ be its quadratic part.

The curvature must act by zero on each matter module, that is, lie in each
matter ideal. The allowed quadratic curvature expressions therefore lie in every
one of these ideals: their intersection gives the bound.

\begin{proposition}\label{prop:integrability}
	Let $\mathfrak p_s\subset R$ be ideals generated by linear expressions in
	the spinor ghosts $\lambda$ and
	containing $I_T$, and let the connection act with charge $\pm1$ on the
	modules $\U\otimes R/\mathfrak p_s$. With $H=\bigcap_s\mathfrak p_s$,
	\[
		\bigl(\sum_i\lambda_i\nabla_i\bigr)^2=0\ \text{on every module}
		\quad\Longleftrightarrow\quad\mathscr F\in\U\otimes H_2,
	\]
	where $\mathscr F$ is taken without its torsion-pair components, which
	vanish by the conventional constraints.
	The largest space of allowed curvature components is $\Delta=H_2/(I_T)_2$,
	and the selected ring is $\OY/\langle\Delta\rangle=R/\langle H_2\rangle$,
	which equals $R/H$ when $H$ is generated in degree two.
\end{proposition}
\begin{proof}
	On each module
	$\bigl(\sum_i\lambda_i\nabla_i\bigr)^2=2i\lp\lbp\nabla_{\pplus}+2i\lm\lbm\nabla_{\mminus}
		\pm i\mathscr F$. The torsion terms vanish in $R/\mathfrak p_s$, because
	$I_T\subset\mathfrak p_s$. The remaining term is multiplication by
	$\pm i\mathscr F$, and applying it to $1\otimes1$ shows that it vanishes
	exactly when $\mathscr F\in\U\otimes\mathfrak p_s$. Since the coefficients
	are taken over a field, intersecting over $s$ commutes with tensoring with
	$\U$, and taking the quadratic part gives the claim.
\end{proof}
This gives an integrability bound. Section~\ref{sec:vectors} constructs
connections attaining it for the examples below. It is only an upper bound:
any smaller set of allowed components, down to the flat ring $\OY$, is equally
compatible with the matter. We always take the largest set, that is, the most
general gauge multiplet compatible with the given matter. This choice
reproduces the known vector multiplets.
The equality $R/\langle H_2\rangle=R/H$ requires
quadratic generation. For a single unconstrained scalar,
$\mathfrak p=(\lp,\lbp,\lm,\lbm)$ and
$R/\langle H_2\rangle=R/\mathfrak p^2\neq k$; the proposition makes no
claim that the selected ring is $R/H$ in such cases. For the two-dimensional matter collections of Section~\ref{sec:matter} the
intersections are generated by quadratic monomials:
\begin{center}\small
	\begin{tabular}{lll}\toprule
		matter, with conjugates                 & $H$                                & allowed curvature $\Delta$              \\\midrule
		chiral                                  & $(\lp,\lm)(\lbp,\lbm)$             & $\lp\lbm,\ \lbp\lm$                     \\
		twisted-chiral                          & $(\lp,\lbm)(\lbp,\lm)$             & $\lp\lm,\ \lbp\lbm$                     \\
		left and right semichiral (four axes)   & $(\lambda_i\lambda_j\mid i\neq j)$ & $\lp\lm,\ \lp\lbm,\ \lbp\lm,\ \lbp\lbm$ \\
		chiral and twisted-chiral (four planes) & $I_T$                              & none                                    \\\bottomrule
	\end{tabular}\end{center}
The first two rows are products of ideals in disjoint variables. A monomial
lies in all four axis ideals exactly when it involves at least two different
ghosts, and in all four plane ideals exactly when it contains $\lp\lbp$ or
$\lm\lbm$. Including conjugates matters for this selection: chiral matter
alone has $H=(\lp,\lm)$, which is linear, and the selected ring is not the
ordinary vector ring. We write
\begin{align*}
	M_V              & =R/(\lp\lbp,\,\lp\lbm,\,\lm\lbp,\,\lm\lbm)=R/(\lp,\lm)(\lbp,\lbm), \\
	M_{\widetilde V} & =R/(\lp\lbp,\,\lp\lm,\,\lbp\lbm,\,\lm\lbm)=R/(\lp,\lbm)(\lbp,\lm), \\
	M_{\mathbb V}    & =R/(\lp\lbp,\,\lm\lbm,\,\lp\lm,\,\lbp\lbm,\,\lp\lbm,\,\lbp\lm).
\end{align*}

In every row $H$ is generated in degree two, so the selected ring is the
coordinate ring of the union of the matter supports: chiral matter selects
two planes, twisted-chiral matter the other two, and the four kinds of
semichiral matter the four axes. Each ideal is closed under conjugation, so
each ring carries a natural real structure. Since the four axes lie in either
pair of planes, the inclusions of ideals give surjections
$M_V\twoheadrightarrow M_{\mathbb V}$ and
$M_{\widetilde V}\twoheadrightarrow M_{\mathbb V}$, whose physical meaning is given
in Section~\ref{sec:reductions}.

The selected ring also has a direct algebraic meaning. The given $R$-action
on $R/\mathfrak p$ factors through $R/J$ exactly when $J\subseteq\mathfrak p$.
Requiring this factorization for all the matter modules gives
$J\subseteq\bigcap_s\mathfrak p_s=H$. Among quadratically generated ideals
contained in $H$, the largest is $J=\langle H_2\rangle$. Being a module over
the selected quotient ring is a stronger condition on matter than being a
module over the CE algebra.

\paragraph{Three and four dimensions.}
The same proof applies. A $3d$ unconstrained scalar and $4d$ chiral matter
with its conjugate give $\Delta=0$, and the selected rings are $\OYd{3}$ and
$\OYd{4}$. Here $\Delta=0$ only means that no spinor--spinor curvature is
allowed; the field strengths sit in the weight-three classes of
Section~\ref{sec:curv-ce}, which a statement about quadrics cannot see.

\paragraph{Complex-linear matter.}
For the higher-order constraint of Section~\ref{sec:linear}, we can check
directly when its ordered covariantization still has a chiral target.
For unit charge in a fixed abelian background, put
$\Ch_\nabla=\ker\bar\nabla_+\cap\ker\bar\nabla_-$. Then $\bar\nabla_+\bar\nabla_-$ maps every charged
unconstrained superfield into $\Ch_\nabla$, and is locally onto, precisely when
\[
	F_{\bpl\bpl}=F_{\bmi\bmi}=F_{\bpl\bmi}=0.
\]
These are the barred-curvature conditions for charged chiral matter,
satisfied by the ordinary vector.

To prove the criterion, write
$\bar\nabla_+^2=\mathcal K_+$, $\bar\nabla_-^2=\mathcal K_-$,
$\{\bar\nabla_+,\bar\nabla_-\}=\mathcal K_{+-}$, where the $\mathcal K$ are multiplication by
the corresponding even curvature coefficients. Then
\[
	\bar\nabla_+(\bar\nabla_+\bar\nabla_-)=\mathcal K_+\bar\nabla_-,\qquad
	\bar\nabla_-(\bar\nabla_+\bar\nabla_-)=\mathcal K_{+-}\bar\nabla_--\mathcal K_-\bar\nabla_+-(\bD_+\mathcal K_-).
\]
Vanishing of the three coefficients suffices. Conversely,
$\{\bar\nabla_+,\bar\theta^+\}=\{\bar\nabla_-,\bar\theta^-\}=1$ and the crossed
anticommutators vanish. Anticommuting the first zero operator with
$\bar\theta^-$ forces $\mathcal K_+=0$; doing the same to the second
with $\bar\theta^+$ and $\bar\theta^-$ forces
$\mathcal K_-=\mathcal K_{+-}=0$. When these vanish, the flat odd partial
connection can be trivialized locally, in coordinates adapted to the barred
distribution, by an invertible even superfield $h$:
$\bar\nabla_+=h\partial_{\bar\theta^+}h^{-1}$,
$\bar\nabla_-=h\partial_{\bar\theta^-}h^{-1}$.
For $\Phi\in\Ch_\nabla$, the superfield
$-h\bar\theta^+\bar\theta^-h^{-1}\Phi$ is a preimage under $\bar\nabla_+\bar\nabla_-$.
Thus the same curvature conditions ensure both the chiral target and local
surjectivity.

The criterion is an instance of Proposition~\ref{prop:integrability}. Since
$c$ is the class of the exact sequence \eqref{eq:linear-splice}, it is
represented by the projection of the complex
$[M_\Phi(-2)\to M_\Phi(-1)\to\C[\lbp]]\simeq k$ onto its first term. This
projection is degreewise onto, so its cone is quasi-isomorphic to the shifted
kernel, and
\begin{equation}
	M_\Sigma\simeq\bigl[M_\Phi(-1)\xrightarrow{\ \lbp\circ r\ }\C[\lbp]\bigr]
	\label{eq:linear-two-term}
\end{equation}
in resolution degrees $-1,0$; its cohomology is $M_\Phi(-2)$ in degree $-1$ and
$k$ in degree $0$, as for $\Cone(c)[-1]$. Both terms, and the map, are
modules over $R/J$ for $J=\langle((\lp,\lm))_2\rangle=(\lp,\lm)(\lp,\lbp,\lm,\lbm)$,
the ring that Proposition~\ref{prop:integrability} selects for the annihilators
$(\lp,\lm)$ of $M_\Phi$ and $(\lp,\lm,\lbm)$ of $\C[\lbp]$; it contains $I_T$.
Hence $\A$ of \eqref{eq:linear-two-term} is a dg module over the commutative
dga $\A(R/J)$, and a connection $\Psi=\sum_i\lambda_i\Gamma_i\in\A(R/J)^1$
twists its differential to $\mathcal Q+i\Psi=\sum_i\lambda_i\nabla_i$. Since
$\Psi^2=0$ for an abelian connection, $(\mathcal Q+i\Psi)^2=i\mathcal Q\Psi$,
which by \eqref{eq:QPsi} is the curvature along the quadrics surviving in
$R/J$, namely $\lbp^2$, $\lbm^2$ and $\lbp\lbm$. So the twisted complex exists
exactly when $F_{\bpl\bpl}=F_{\bmi\bmi}=F_{\bpl\bmi}=0$. Only the barred
ghosts act nonzero on the two modules, so only $\Gamma_{\bpl},\Gamma_{\bmi}$
enter, and conjugation by the trivialization $h$ above identifies the twisted
complex with the untwisted one; its cohomology is therefore
$\ker\bar\nabla_+\bar\nabla_-$, in agreement with the superspace criterion. The coupling of complex-linear
matter to a fixed abelian connection is thus a module-level statement over
$R/J$, with the same selection principle as for first-order matter.

The criterion has a direct reading in terms of the first-order system
\eqref{eq:first-order}. Its covariantization
\[
	\mathcal L_\nabla=\begin{pmatrix}\bar\nabla_+ & -1\\ 0 & \bar\nabla_-\end{pmatrix}\colon\
	\U\oplus\Pi\ker\bar\nabla_+\longrightarrow\Pi\ker\bar\nabla_+\oplus\Ch_\nabla
\]
is well defined exactly under the three conditions above: $\bar\nabla_+$
maps into $\ker\bar\nabla_+$ when $\mathcal K_+=0$, and for
$\bar\nabla_+S=0$ one has $\bar\nabla_+\bar\nabla_-S=\mathcal K_{+-}S$ and
$\bar\nabla_-^2S=\mathcal K_-S$, so $\bar\nabla_-$ maps $\ker\bar\nabla_+$ into
$\Ch_\nabla$ when $\mathcal K_{+-}=\mathcal K_-=0$. The row and column
operations \eqref{eq:elimination} then apply verbatim and eliminate the
auxiliary field, with Schur complement
$\bar\nabla_-\bar\nabla_+=-\bar\nabla_+\bar\nabla_-$. The curvature
conditions are thus precisely those under which the elimination of
Section~\ref{sec:linear} survives covariantization.

\paragraph{Chiral and twisted-chiral matter together.}
By the last row of the table, a single connection acting on chiral and
twisted-chiral matter and their conjugates can carry no spinor curvature,
and it is then flat. Indeed, with the conventional constraints
$F_{+\bpl}=F_{-\bmi}=0$ all spinor anticommutators of the covariant
derivatives vanish except $\{\nabla_+,\bar\nabla_+\}=2i\nabla_{\pplus}$ and
$\{\nabla_-,\bar\nabla_-\}=2i\nabla_{\mminus}$. The Jacobi identity
$2[X,\{X,Y\}]=[\{X,X\},Y]$ for odd operators $X,Y$, and its analog for three different odd
operators show that every spinor derivative commutes with $\nabla_{\pmpm}$,
and then
$2i[\nabla_{\pplus},\nabla_{\mminus}]=\{[\nabla_{\pplus},\nabla_-],\bar\nabla_-\}
	+\{\nabla_-,[\nabla_{\pplus},\bar\nabla_-]\}=0$. All curvature vanishes, and
the super Poincar\'e lemma gives, locally,
\begin{equation}
	\Gamma=d\Lambda_0,\qquad\nabla=e^{-i\Lambda_0}De^{i\Lambda_0},
	\label{eq:flatness}
\end{equation}
for a not necessarily supersymmetric primitive $\Lambda_0$. The large vector
multiplet avoids this obstruction by using two connections with a common
gauge parameter \cite{NewVectors,Nonabelian} (Section~\ref{sec:lvm}).

\subsection{The two-dimensional gauge rings and their field strengths}\label{sec:rings}
In two dimensions the spinor curvature components correspond to quadrics.
Beyond torsion there are eight: the four squares $\lambda_i^2$ and the four mixed quadrics
$\lp\lm$, $\lbp\lbm$, $\lp\lbm$, $\lbp\lm$. For the matter collections above the
squares always survive: each ghost survives in some matter quotient.
Thus the diagonal curvature components satisfy $F_{ii}=0$. For the real
connections considered here, we require the defining ideal to be preserved
by conjugation $\lambda_\pm\leftrightarrow\bar\lambda_\pm$,
which exchanges $\lp\lm\leftrightarrow\lbp\lbm$ and
$\lp\lbm\leftrightarrow\lbp\lm$. Under these conditions the
conjugation-invariant monomial choices for the two conjugate pairs are the
four rings below; we do not classify non-monomial real subspaces.
\begin{center}\small
	\begin{tabular}{@{}lllll@{}}\toprule
		ring               & imposed besides $F_{ii}=0$                & allowed beyond torsion & Betti       & multiplet         \\\midrule
		$\OY$              & $F_{+-},F_{\bpl\bmi},F_{+\bmi},F_{\bpl-}$ & none                   & $(1,2,1)$   & flat (de Rham)    \\
		$M_V$              & $F_{+-},F_{\bpl\bmi}$                     & $F_{+\bmi},F_{\bpl-}$  & $(1,4,4,1)$ & vector, $4+4$     \\
		$M_{\widetilde V}$ & $F_{+\bmi},F_{\bpl-}$                     & $F_{+-},F_{\bpl\bmi}$  & $(1,4,4,1)$ & twisted, $4+4$    \\
		$M_{\mathbb V}$    & none                                      & all four               & $(1,6,8,3)$ & semichiral, $8+8$ \\\bottomrule
	\end{tabular}\end{center}
The Betti numbers are computed over $R$: for $\OY$ they come from the
Koszul resolution on its two defining quadrics, and the remaining rows
follow from Appendix~\ref{app:betti}. Component counts of real multiplets,
here and below, are real counts at generic momentum after the gauge
quotient; they equal the complex dimensions read off from $\Tor$ of the
complexified complex. Counts of complex matter multiplets (chiral,
semichiral, complex-linear) are complex counts. For the ordinary
vector the constraints are the standard ones,
$\{\nabla_+,\nabla_-\}=\{\bar\nabla_+,\bar\nabla_-\}=\nabla_i^2=0$, the
conditions for covariantly chiral and antichiral superfields; the twisted
vector is its partial conjugate, and the semichiral vector imposes only
$F_{ii}=0$. The choice among the four is made by the matter
(Section~\ref{sec:selection}), with one ring for each matter collection: the
four kinds of semichiral superfield all couple to the same semichiral vector.
The large vector multiplet does not appear in the list. It gauges chiral and
twisted-chiral matter at once, so by \eqref{eq:flatness} it cannot
come from a single connection and a quotient ring; it is treated in
Section~\ref{sec:lvm}.

The allowed curvature components span the ideal $\mathfrak a=\ker(\OY\to M)$.
For these rings, this ideal splits into shifted quotients of $R$ by linear
monomials. Their reconstruction therefore gives constrained superfields by
\eqref{eq:linsupp}.
The exact sequence
\[
	0\longrightarrow\mathfrak a\longrightarrow\OY\longrightarrow M\longrightarrow0
\]
has a connecting morphism that reconstructs the curvature map, as the
following proposition shows.

\begin{proposition}\label{prop:strengths}
	Let $M$ be one of $M_V$, $M_{\widetilde V}$, $M_{\mathbb V}$, and let
	$\mathfrak a=\ker(\OY\to M)$ be the ideal spanned by the allowed curvature components.
	\begin{enumerate}[label=(\roman*)]
		\item $\mathfrak a\cong\bigoplus_a\C[Z_a](-2)$, where $Z_a$ runs over the coordinate
		      planes whose two variables $x_a,y_a$ satisfy $x_ay_a\in\mathfrak a$, and the summand
		      is generated by $x_ay_a$.
		\item Locally, $\A(M)[1]$ has cohomology $\C$, the constants, in degree $-1$,
		      and in degree $0$ the map $\Psi\mapsto(F_{x_ay_a})_a$ is an isomorphism
		      \[
			      H^0\bigl(\A(M)[1]\bigr)\;\cong\;\bigoplus_a
			      \bigl\{G\in\U:\ D_iG=0\text{ for all }\lambda_i\text{ on }Z_a\bigr\}.
		      \]
		      There is no other cohomology.
	\end{enumerate}
\end{proposition}
\begin{proof}
	(i) In $\OY$ a monomial is nonzero exactly when it contains no torsion pair,
	that is, when its variables lie on one of the four planes. A monomial of $\mathfrak a$
	is therefore divisible by $x_ay_a$ for exactly one plane $Z_a$, and it is
	$x_ay_a$ times a monomial in $x_a,y_a$. Every variable off $Z_a$ is the
	torsion partner of $x_a$ or of $y_a$ and annihilates $x_ay_a$, so the submodule
	generated by $x_ay_a$ is $x_ay_a\C[Z_a]\cong\C[Z_a](-2)$, and the sum over
	$a$ is direct because the monomial supports are disjoint.

	(ii) Reconstruct the exact sequence above, using exactness of $\A$. Locally
	$\A(\OY)\simeq\A(\CE{2})$ has cohomology $\C$ in degree zero, since the de Rham
	complex resolves the locally constant functions \cite[Lemma~3.8]{Derived}.
	By (i) and \eqref{eq:linsupp}, $\A(\mathfrak a)$ has cohomology
	$\bigoplus_a\{G:D_iG=0\text{ on }Z_a\}$ in degree two and nowhere else. The
	long exact sequence gives $H^0(\A(M))=\C$ and
	$H^1(\A(M))\cong H^2(\A(\mathfrak a))$, and nothing else. The connecting map lifts a
	cocycle $\Psi\in\U\otimes M_1=\U\otimes (\OY)_1$ to $\OY$ and applies
	$\mathcal Q$; by \eqref{eq:QPsi} the result lies in $\U\otimes\mathfrak a_2$, and its
	component along $x_ay_a$ is $F_{x_ay_a}$.
\end{proof}

For the three rings:
\begin{itemize}
	\item $M_V$: $\mathfrak a$ is spanned over the twisted-chiral plane by $\lbp\lm$
	      and over its conjugate by $\lp\lbm$. The strengths are $F_{\bpl-}$, twisted
	      chiral, and its conjugate $F_{+\bmi}$: the field strength $\sigma$ of the
	      $2d$ vector multiplet and $\bar\sigma$.
	\item $M_{\widetilde V}$: the strengths are $F_{\bpl\bmi}$, chiral, and $F_{+-}$,
	      antichiral.
	\item $M_{\mathbb V}$: all four mixed components are allowed, and the strengths
	      are one chiral, one twisted-chiral and their conjugates
	      (Section~\ref{sec:svm}).
\end{itemize}
In each case connections modulo gauge are the same as field strengths, and
only the constant gauge transformations act trivially. This is the connection-form analog of
\eqref{eq:gauge-exact}; Section~\ref{sec:vectors} shows that it agrees
with the potential form.

\paragraph{Relation to the ordinary and twisted vectors.}
The ideals of $M_V$ and $M_{\widetilde V}$ intersect in $I_T$ and sum to the
ideal of $M_{\mathbb V}$. Indeed, a monomial without a torsion pair lies on one
plane; it lies in the ideal of $M_V$ exactly when its plane is twisted chiral
or its conjugate and it contains both variables, and in the ideal of
$M_{\widetilde V}$ exactly when its plane is chiral or antichiral and it
contains both variables, and no monomial does both. Equivalently,
$\mathfrak a_{\mathbb V}=\mathfrak a_V\oplus\mathfrak a_{\widetilde V}$, and
\begin{equation}
	0\to \OY\to M_V\oplus M_{\widetilde V}\to M_{\mathbb V}\to0,\qquad
	s\mapsto(s,s),\quad(a,b)\mapsto a|_{\rm axes}-b|_{\rm axes},
	\label{eq:mv}
\end{equation}
is exact. Its reconstruction, in which $\A(\OY)[1]$ contributes only the
constants in degree $-1$, gives locally
$H^0(\A(M_{\mathbb V})[1])\cong H^0(\A(M_V)[1])\oplus
	H^0(\A(M_{\widetilde V})[1])$: the gauge-invariant content of the semichiral
vector is that of an ordinary and a twisted vector together. The constant
gauge transformations $\C\oplus\C$ of the middle term map onto the constants
$\C$ of the semichiral vector, with the diagonal as kernel.

\paragraph{Two- versus four-dimensional field strengths.}
The rings $M_V$ and $\OYd{4}$ are isomorphic: $\lambda^\alpha\mapsto(\lp,\lm)$,
$\bar\lambda^{\dot\alpha}\mapsto(\lbp,\lbm)$ identifies the defining
ideals. The multiplets differ because the torsion ideal sits differently
inside the defining ideal $J$: in four dimensions all four generators are
torsion, while in two dimensions only $\lp\lbp$ and $\lm\lbm$ are, and the
other two correspond to $\sigma$ and $\bar\sigma$. Under dimensional
reduction the transverse translations drop out of the anticommutators, so
$\lambda^1\bar\lambda^{\dot2}$ and $\lambda^2\bar\lambda^{\dot1}$ become
allowed curvature components and the transverse components of the gauge field
become $\sigma$.

\section{Potentials and gauge complexes}\label{sec:vectors}
In superspace a vector multiplet is usually represented by unconstrained
potentials with gauge transformations, such as $V$ with
$\delta V=i(\bar\Lambda-\Lambda)$ where $\Lambda$ is a chiral superfield.
In this section we recover potentials and gauge
transformations from restriction sequences for the modules.
Their reconstruction gives the gauge complexes of Section~\ref{sec:gauge-cone},
with potentials appearing as differences of partial potentials on the pieces
of the support. The three-dimensional spinor potential is obtained directly
from its module, without a restriction sequence. All constructions in this
section are abelian.

\subsection{Partial potentials and restriction sequences}\label{sec:cones}

\paragraph{An example: the four-dimensional vector multiplet.}
The ring
$\OYd{4}=\C[\lambda^\alpha,\bar\lambda^{\dot\alpha}]/(\lambda^\alpha\bar\lambda^{\dot\alpha})$
is the coordinate ring of two planes meeting at the origin: the plane
$\bar\lambda=0$, with variables $\lambda^\alpha$, and the plane $\lambda=0$,
with variables $\bar\lambda^{\dot\alpha}$. We call these the two
\emph{pieces} of its support. Monomials in the variables of one piece, such
as $\lambda^\alpha\lambda^\beta$, survive in the ring and give the
constraints $F_{\alpha\beta}=F_{\dot\alpha\dot\beta}=0$
(Section~\ref{sec:curv-ce}): restricted to each piece, the connection is
flat. The mixed monomials $\lambda^\alpha\bar\lambda^{\dot\alpha}$ vanish;
they are the torsion quadrics. A flat connection is locally pure gauge,
so
\[
	\Gamma_\alpha=D_\alpha\Omega,\qquad
	\bar\Gamma_{\dot\alpha}=\bD_{\dot\alpha}\bar\Omega
\]
for two even superfields $\Omega$ and $\bar\Omega$, one for each
piece; we call them \emph{partial potentials}. This is the familiar
derivation of the potential \cite{Superspace}, in which $\Omega$ and
$\bar\Omega$ define the antichiral and chiral representations. Each
partial potential is fixed only up to the gauge parameter $\kappa$ of the
connection, $\delta\Gamma_A=D_A\kappa$, and up to superfields annihilated by the
derivatives of its piece:
\[
	\Omega\mapsto\Omega+\kappa+\bar\Lambda',\qquad
	\bar\Omega\mapsto\bar\Omega+\kappa+\Lambda,
\]
with $\bar\Lambda'$ antichiral ($D_\alpha\bar\Lambda'=0$) and $\Lambda$
chiral ($\bD_{\dot\alpha}\Lambda=0$). The common $\kappa$ cancels in the
difference, and $V=i(\Omega-\bar\Omega)$ transforms as
$\delta V=i(\bar\Lambda'-\Lambda)$; on the real form, $\bar\Lambda'=\bar\Lambda$,
this is the usual $\delta V=i(\bar\Lambda-\Lambda)$. So the potential of
the $4d$ vector multiplet is the difference of the two partial potentials,
and its gauge parameters are the constrained superfields of the two pieces.

\paragraph{The general statement.}
As in four dimensions, the same structure appears for the two-dimensional gauge rings $M_V$ and
$M_{\widetilde V}$, each the union of two planes, and $M_{\mathbb V}$, the
union of the four coordinate axes. In all these cases the support is a union
$Z=Z_1\cup\cdots\cup Z_n$ of coordinate subspaces of the nilpotence variety,
the \emph{pieces}, such that every ghost $\lambda_i$ is a coordinate on
exactly one piece; we say that $\lambda_i$ \emph{lies on} that piece.
Equivalently, the pieces meet only at the origin and together use all the
ghosts. The surviving quadrics in $M=R/I(Z)$ impose $F_{ij}=0$ whenever
$\lambda_i$ and $\lambda_j$ lie on the same piece, $i=j$ included.
Mixed quadrics impose no additional spinorial curvature equation; those
corresponding to torsion are instead handled by the conventional constraints
of Section~\ref{sec:curv-ce}. For $\OYd{4}$ all mixed quadrics are torsion.

On each piece the derivatives $D_i$ anticommute and square to zero, and in
coordinates adapted to the piece they are partial derivatives along odd
coordinates. The constraints say that the odd one-form with components
$\Gamma_i$ is closed in these directions, so by the odd Poincar\'e lemma
with parameters it has a primitive on any coordinate neighborhood:
\begin{equation}
	\Gamma_i=D_i\Omega_a\quad\text{for }\lambda_i\text{ on }Z_a,\qquad
	\Omega_a\mapsto\Omega_a+\kappa+\eta_a,\qquad
	\delta(\Omega_a-\Omega_b)=\eta_a-\eta_b .
	\label{eq:frames}
\end{equation}
There is one even superfield $\Omega_a$ for each piece, the
\emph{partial potential}. It is unique up to superfields $\eta_a$
annihilated by all $D_i$ of the piece, which are the constrained superfields of
the matter module $\C[Z_a]$ (see \eqref{eq:linsupp}), and the gauge
transformation $\delta\Gamma_i=D_i\kappa$ shifts every partial potential by
$\kappa$. Since every ghost lies on exactly one piece, the partial
potentials determine the connection, and their differences do not depend on
$\kappa$. Changing the local primitives changes their differences only by
the displayed gauge transformations.
Choosing $Z_n$ as a reference piece gives $n-1$ relative potentials
\[
	P_a=\Omega_a-\Omega_n,\qquad
	\delta P_a=\eta_a-\eta_n,\qquad 1\leq a<n.
\]
A different reference piece gives an equivalent presentation.

\paragraph{Restriction sequences.}
On the module side the partial potentials come from a short exact
sequence. A function on a union of $n$ pieces $Z_a$ meeting only at the
origin is a tuple of functions on the pieces with equal values there:
\begin{equation}
	0\to M\to\bigoplus_a\C[Z_a]\xrightarrow{\ \mathrm{ev}\ }k^n/k\to0,
	\qquad
	\mathrm{ev}(f)=\bigl(f_a(0)-f_n(0)\bigr)_{a<n},
	\label{eq:restriction}
\end{equation}
where the reference piece $Z_n$ identifies $k^n/k\cong k^{n-1}$; so
$M[1]\simeq\Cone(\mathrm{ev})$. By \eqref{eq:linsupp}, $\C[Z_a]$
reconstructs to the gauge parameters of piece $a$,
\[
	\G_a=\{\eta\in\U:\ D_i\eta=0\text{ for all }\lambda_i\text{ on }Z_a\},
\]
and $\A(\mathrm{ev})$ becomes their differences
$g_{\rm pot}(\eta)=(\eta_a-\eta_n)_{a<n}$. Hence
\[
	\A(M)[1]\simeq\Bigl[\bigoplus_a\G_a
		\xrightarrow{\ g_{\rm pot}\ }\U^{\,n-1}\Bigr]_{-1,0}:
\]
the target holds the relative potentials $P_a$, and the map is their gauge
transformation. For $\OYd{4}$ the pieces reconstruct to antichiral and
chiral superfields, and this is $[\bar\Ch\oplus\Ch\to\U]_{-1,0}$,
$(\bar\Lambda',\Lambda)\mapsto\bar\Lambda'-\Lambda$: the potential $-iV$ of
the example, with the constant gauge transformations in degree $-1$. Real
forms are identified afterwards, case by case.

\begin{theorem}\label{thm:vectors}
	For $M=\OYd{4}$, $M_V$, $M_{\widetilde V}$, $M_{\mathbb V}$,
	\[
		\A(M)[1]\simeq\Bigl[\bigoplus_a\G_a
			\xrightarrow{\ g_{\rm pot}\ }\U^{\,n-1}\Bigr]_{-1,0},
	\]
	a local quasi-isomorphism with the two-term complex of differences of partial
	potentials with their gauge transformations: the $4d$ vector multiplet, the ordinary and twisted $2d$
	vector multiplets, and the semichiral vector multiplet.
	In every case the constant gauge transformations form the locally constant
	line in degree $-1$.
\end{theorem}
\begin{proof}
	The sequence \eqref{eq:restriction} is exact: an element of $M$ is a
	constant plus terms of positive degree on each piece, because monomials
	mixing two pieces vanish, and a tuple lies in the image exactly when its
	values at the origin agree; $\mathrm{ev}$ is clearly onto. Reconstruction preserves
	cones, by \eqref{eq:linsupp} we can identify each
	$\A(\C[Z_a])$ with $\G_a$, and these inclusions commute
	with $\A(\mathrm{ev})$ because $\mathrm{ev}$ only evaluates ghosts at zero. A parameter with zero
	gauge variation has $\eta_a=\eta_b$ for all $a,b$; it is then
	annihilated by every spinor derivative, hence constant. For $\OYd{4}$ the statement
	concerns $\OYd{4}$-modules restricted along $\CE{4}\to \OYd{4}$
	(Section~\ref{sec:setting}).
\end{proof}
For $M_V$, $M_{\widetilde V}$ and $M_{\mathbb V}$, combining
Theorem~\ref{thm:vectors} with Proposition~\ref{prop:strengths} gives the
exact sequence \eqref{eq:gauge-exact}: potentials modulo gauge are field
strengths.

\paragraph{The fiber-product principle.}
In each case of Theorem~\ref{thm:vectors} the pieces $Z_a$ are the supports of
the matter the multiplet gauges, together with their conjugates, and $M$ is
the fiber product $\C[Z_1]\times_k\cdots\times_k\C[Z_n]$, the kernel of
$\mathrm{ev}$ in \eqref{eq:restriction}. This is the organizing principle of
the gauge multiplets in this paper: \emph{the gauge module is the fiber
product over the origin of the coordinate rings of the matter supports}, and
its gauge parameters are the matter superfields of the pieces. When the
pieces meet only at the origin, as here, the fiber product equals the
quotient ring $R/\langle H_2\rangle$ selected by
Proposition~\ref{prop:integrability}, and the curvature description of
Section~\ref{sec:gauge} applies. When they overlap, the fiber product is not a
quotient of $\OY$ and a single connection is too restrictive; this is the
large vector multiplet of Section~\ref{sec:lvm}.

\paragraph{Agreement with the four-dimensional resolution.}
For $\OYd{4}$ the restriction sequence recovers both the module and the
minimal free resolution of \cite[Section~3.8]{Perspectives}, and shows
where the quadratic relations come from. Write
\[
	R=\C[\lambda^1,\lambda^2,\bar\lambda^{\dot1},\bar\lambda^{\dot2}],\qquad
	\mathfrak m=(\lambda^1,\lambda^2),\qquad
	\bar{\mathfrak m}=(\bar\lambda^{\dot1},\bar\lambda^{\dot2}).
\]
For $n=2$, \eqref{eq:restriction} reads
\[
	0\longrightarrow M\longrightarrow
	R/\mathfrak m\oplus R/\bar{\mathfrak m}
	\xrightarrow{\ \mathrm{ev}\ }\C\longrightarrow0,
	\qquad \mathrm{ev}(f,g)=f(0)-g(0),
\]
and since $\mathfrak m\cap\bar{\mathfrak m}=\mathfrak m\bar{\mathfrak m}$,
its kernel is $M=R/(\lambda^\alpha\bar\lambda^{\dot\alpha})$, the module of
\cite{Perspectives}.

To obtain the resolution, resolve the two planes and their intersection by
the Koszul complexes
$R\otimes\Lambda(\bar\xi_{\dot\alpha})$, $R\otimes\Lambda(\xi_\alpha)$ and
$R\otimes\Lambda(\xi_\alpha,\bar\xi_{\dot\alpha})$, with
$d\xi_\alpha=\lambda^\alpha$ and $d\bar\xi_{\dot\alpha}=\bar\lambda^{\dot\alpha}$.
The map $\mathrm{ev}$ lifts to the difference of the inclusions of exterior
algebras. Its cone $C$, shifted so that $C\simeq M$, is $R\otimes\mathcal B$ with
\[
	\mathcal B=\Lambda(\bar\xi)\oplus\Lambda(\xi)\xrightarrow{\ d_0\ }\Lambda(\xi,\bar\xi),
\]
and its differential is $d=d_0+d_1$, where $d_1$ consists of the Koszul
differentials, with linear entries. In exterior degree $p$, $d_0$ sends
$(a,b)\in\Lambda^p(\bar\xi)\oplus\Lambda^p(\xi)$ to $a-b$. For $p=0$ it is
onto, with kernel spanned by $\mathbf 1=(1,1)$, the sum of the units of the
two planes. For $p\ge1$ it is injective, with image the monomials lying on a
single plane, since $\Lambda^p(\xi)\cap\Lambda^p(\bar\xi)=0$. Hence
$H^\bullet(\mathcal B,d_0)$ is spanned by the common unit and the mixed monomials,
\[
	H^\bullet(\mathcal B,d_0)=
	\C\,\mathbf 1\oplus
	\langle\xi_\alpha\bar\xi_{\dot\alpha}\rangle\oplus
	\langle\xi_\alpha\bar\xi_1\bar\xi_2,\ \xi_1\xi_2\bar\xi_{\dot\alpha}\rangle\oplus
	\langle\xi_1\xi_2\bar\xi_1\bar\xi_2\rangle,
\]
of ranks $(1,4,4,1)$ in resolution degrees $0,-1,-2,-3$. A $\C$-linear splitting of $\mathcal B$ gives maps
$\iota,\pi$ and a homotopy $h$ with $d_0h+hd_0=1-\iota\pi$, and the
perturbation lemma transfers $d_1$ to
\[
	\delta=\pi d_1\iota-\pi d_1hd_1\iota+\pi d_1hd_1hd_1\iota-\cdots
\]
on $R\otimes H^\bullet(\mathcal B,d_0)$. The series terminates, because $h$ preserves and
$d_1$ lowers the exterior degree. The result is a free complex
quasi-isomorphic to $C$, hence a resolution of $M$, and it is minimal
because all entries of $\delta$ lie in $(\lambda,\bar\lambda)$:
\begin{equation}
	0\longrightarrow R(-4)\longrightarrow R(-3)^4\longrightarrow R(-2)^4
	\longrightarrow R\longrightarrow M\longrightarrow0,
	\label{eq:oy4-res}
\end{equation}
with one generator for each basis element of $H^\bullet(\mathcal B,d_0)$ above.

Only one kind of entry needs more than the first term. On the cubic and
quartic mixed monomials, $\delta$ is the Koszul differential on
$\Lambda(\xi,\bar\xi)$ with the terms lying on a single plane discarded by
$\pi$; these are the linear maps of \eqref{eq:oy4-res}. On a quadratic
mixed monomial the first term vanishes, because
\[
	d_1(\xi_\alpha\bar\xi_{\dot\alpha})
	=\lambda^\alpha\bar\xi_{\dot\alpha}-\bar\lambda^{\dot\alpha}\xi_\alpha
\]
has one piece on each plane. The second term lifts these pieces to the two
planes and applies their Koszul differentials. Each gives
$\lambda^\alpha\bar\lambda^{\dot\alpha}$ times the unit of its plane, so
together they give the common unit:
\[
	\delta(\xi_\alpha\bar\xi_{\dot\alpha})
	=-\pi d_1hd_1\iota(\xi_\alpha\bar\xi_{\dot\alpha})
	=\lambda^\alpha\bar\lambda^{\dot\alpha}\,\mathbf 1.
\]
Thus the linear syzygies come from the Koszul complex of the intersection
point, and each quadratic generator of the ideal is the obstruction to
splitting a mixed monomial between the two planes. This reproduces the
minimal resolution of \cite{Perspectives}.

\paragraph{The three-dimensional vector.}
The ring $\OYd{3}$ has only weights zero and one, so directly
\[
	\A(\OYd{3})[1]=\bigl[\U\xrightarrow{\kappa\mapsto\lambda^\alpha D_\alpha\kappa}
		\U\otimes\langle\lambda^1,\lambda^2\rangle\bigr]_{-1,0}.
\]
The coefficient of $\lambda^\alpha$ is an odd spinor $\Gamma_\alpha$ with
$\delta\Gamma_\alpha=D_\alpha\kappa$, and since all quadratic ghost products
vanish there is no further constraint: this is the spinor-potential form
of the $3d$ vector \cite{Superspace,KT}. If both $D_\alpha\kappa$ vanish,
their anticommutator forces all spacetime derivatives of $\kappa$ to
vanish, so again the constant gauge transformations form the line $\C$ in degree $-1$.
In Wess--Zumino gauge the three-dimensional vector contains $A_a$ and a
Majorana gaugino, with no auxiliary field, giving $2+2$ real components
after the gauge quotient at generic momentum.

\subsection{The ordinary and twisted vectors}\label{sec:ordinary}
Chiral matter with its conjugate selects the ring $M_V$
(Section~\ref{sec:selection}): the curvature is restricted to the
components $F_{+\bmi}$ and $F_{\bpl-}$, while all others vanish. Under
$\lambda^\alpha\mapsto(\lp,\lm)$, $\bar\lambda^{\dot\alpha}\mapsto(\lbp,\lbm)$
this ring is isomorphic to $\OYd{4}$ (Section~\ref{sec:rings}), so the
solution is the four-dimensional one of Section~\ref{sec:cones}, with
$\alpha,\dot\alpha$ replaced by $\pm,\bar\pm$: partial potentials
$\Omega$ and $\bar\Omega$ on the unbarred and barred planes, and,
on the real form where $\bar\Omega$ is the complex conjugate of $\Omega$, the real potential
$V=i(\Omega-\bar\Omega)$ with $\delta V=i(\bar\Lambda-\Lambda)$. The
allowed curvature is the field strength:
\[
	F_{\bpl-}=\bD_+D_-(\Omega-\bar\Omega)=-i\,\sigma,\qquad
	\sigma=\bD_+D_-V,
\]
twisted chiral because $\bD_+^2=D_-^2=\{\bD_+,D_-\}=0$, and $F_{+\bmi}$ is
its conjugate. This is the familiar field strength $\sigma$ of the $2d$
vector multiplet, in agreement with Proposition~\ref{prop:strengths}. The
multiplet has $4+4$ real components: $A_{\pmpm}$, a complex scalar, four
real fermions and a real auxiliary field.

Twisted-chiral matter selects $M_{\widetilde V}$, the partial conjugate. The
real potential $\widetilde V$ is the difference of partial potentials on the
two twisted-chiral planes, with a twisted-chiral gauge parameter, and the
field strength $\bD_+\bD_-\widetilde V$ is chiral. Twisted-chiral matter has
no four-dimensional counterpart.

\subsection{The semichiral vector}\label{sec:svm}
Off-shell $(2,2)$ sigma models on generalized K\"ahler targets use chiral,
twisted-chiral and semichiral superfields \cite{GHR,BLR,LRvUZ}, and gauging
their isometries requires new vector multiplets: semichiral parameters give
the semichiral vector multiplet (SVM) \cite{NewVectors,GatesMerrell}, and
chiral and twisted-chiral parameters together the large vector multiplet
(LVM) \cite{NewVectors}. Both have three real potentials, which combine into
two complex ones, $\mathbb V,\mathbb W$ for the SVM and $\mathcal V,\mathcal W$
for the LVM \cite{NewVectors,Nonabelian} (\cite{Nonabelian} writes
$V_L,V_R$ for $\mathcal V,\mathcal W$). The SVM has a chiral and a
twisted-chiral field strength and reduces to the ordinary or the twisted
vector when one of them vanishes \cite{GatesMerrell,SVM}; it gives gauged
linear sigma models with torsion \cite{SVM}. The LVM has semichiral spinor
field strengths, and its action was found in \cite{Ryb}. The SVM is selected by semichiral matter and
comes from a quotient ring, like the ordinary vectors. Simultaneous chiral
and twisted-chiral matter constraints select a flat common connection,
not the LVM; the latter comes from a fiber-product algebra on four
overlapping planes (Section~\ref{sec:lvm}).

Left and right semichiral matter with its conjugates selects the ring
$M_{\mathbb V}$ (Section~\ref{sec:selection}): only $F_{ii}=0$ is imposed,
and all four mixed curvatures $F_{+-}$, $F_{\bpl\bmi}$, $F_{+\bmi}$ and
$F_{\bpl-}$ are allowed. Its minimal free resolution over $R$,
\[
	0\to R(-4)^3\to R(-3)^8\to R(-2)^6\to R\to M_{\mathbb V}\to0,
\]
has the six quadrics as generators, eight linear first syzygies and three
linear second syzygies; the Betti numbers are derived in
Appendix~\ref{app:betti}. After the shift by
one, the constant generator gives the scalar gauge parameter in degree $-1$
and all other classes lie in degree zero: six bosonic connection and scalar
components, eight fermions and three bosonic auxiliary fields. At generic
momentum this gives $(6+3-1)+8=8+8$ real components.

\paragraph{Partial potentials.}
Solving $F_{ii}=0$ gives four partial potentials, $\Gamma_i=D_i\Omega_i$
(see \eqref{eq:frames}), one for each axis of $M_{\mathbb V}$. Each is
fixed up to a superfield annihilated by one spinor derivative:
$\eta_{\bpl}=\mathbb X_L$ and $\eta_{\bmi}=\mathbb X_R$ are left and right
semichiral, and $\eta_+=\bar{\mathbb X}_L$ and $\eta_-=\bar{\mathbb X}_R$ satisfy
the conjugate constraints. Here and below, as usual in superspace, a field
and its barred partner are treated as independent in the complexified
complex; the reality condition then identifies the barred field with the
complex conjugate. Four complex partial potentials modulo a common shift
have three independent complex differences; imposing reality gives the
three real potentials. With
\begin{equation}
	\mathbb V=\Omega_{\bpl}-\Omega_{\bmi},\qquad
	\mathbb W=\Omega_{\bpl}-\Omega_-,\qquad
	\bar{\mathbb V}=\Omega_+-\Omega_-,\qquad
	\bar{\mathbb W}=\Omega_+-\Omega_{\bmi},\label{eq:svm-frames}
\end{equation}
one has $\mathbb V+\bar{\mathbb V}=\mathbb W+\bar{\mathbb W}$ identically, and
\begin{equation}
	\delta \mathbb V=\mathbb X_L-\mathbb X_R,\qquad\delta \mathbb W=\mathbb X_L-\bar{\mathbb X}_R,\qquad
	\delta \bar{\mathbb V}=\bar{\mathbb X}_L-\bar{\mathbb X}_R,\qquad
	\delta \bar{\mathbb W}=\bar{\mathbb X}_L-\mathbb X_R.\label{eq:complex-gauge}
\end{equation}
The reality condition $\Omega_{\bar\imath}=\overline{\Omega_i}$ makes
$\bar{\mathbb V}$, $\bar{\mathbb W}$, $\bar{\mathbb X}_L$ and $\bar{\mathbb X}_R$
the complex conjugates of $\mathbb V$, $\mathbb W$, $\mathbb X_L$ and $\mathbb X_R$, and
\eqref{eq:complex-gauge} becomes the SVM of \cite[(2.10)--(2.11)]{Nonabelian};
relative to \cite[(2.15)]{NewVectors} the dictionary is
$\mathbb X_L=\Lambda_L$, $\mathbb X_R=-\Lambda_R$.

\paragraph{Field strengths.}
Bars on the strengths below label conjugate sectors, not literal complex
conjugation: on the real form our derivative ordering gives
$\bar f=-f^*$ and $\bar{\tilde f}=-\tilde f^*$, where ${}^*$ denotes
complex conjugation.
For two directions that do not form a torsion pair, $D_jD_i=-D_iD_j$, so
$F_{ij}=D_iD_j(\Omega_j-\Omega_i)$ depends only on a difference of partial
potentials. With \eqref{eq:svm-frames} the four allowed curvature
components are $F_{\bpl\bmi}=-f$, $F_{\bpl-}=-\tilde f$, $F_{+-}=-\bar f$ and
$F_{+\bmi}=-\bar{\tilde f}$, with
\begin{equation}
	f=\bD_+\bD_-\mathbb V,\qquad\tilde f=\bD_+D_-\mathbb W,\qquad
	\bar f=D_+D_-\bar{\mathbb V},\qquad
	\bar{\tilde f}=D_+\bD_-\bar{\mathbb W}.\label{eq:strength}
\end{equation}
They are gauge invariant, because $D_iD_j$ annihilates the ambiguities
$\eta_i$ and $\eta_j$ of the two partial potentials involved,
and they are constrained superfields, because the spinor derivatives square
to zero:
\begin{equation}
	\bD_\pm f=0,\qquad\bD_+\tilde f=D_-\tilde f=0,\qquad
	D_\pm \bar f=0,\qquad D_+\bar{\tilde f}=\bD_-\bar{\tilde f}=0.
	\label{eq:strength-target}
\end{equation}
So $f$ is chiral and $\tilde f$ twisted chiral. The conjugation signs stated
above come from conjugating products of odd derivatives. In
\cite[(2.16)]{NewVectors} the conjugate strengths are defined with an extra
sign, $-D_+D_-\bar{\mathbb V}$ and $-D_+\bD_-\bar{\mathbb W}$, and are the
complex conjugates of $f$ and $\tilde f$.

By Proposition~\ref{prop:strengths} these four constrained superfields identify
the degree-zero cohomology of the gauge complex; the constant gauge
transformations remain in degree $-1$. Locally,
\begin{equation}
	0\to\C\to\G\xrightarrow{\ g_{\rm pot}\ }\Pot
	\xrightarrow{\ \str\ }
	\bigl\{(f,\tilde f,\bar f,\bar{\tilde f})\ \text{obeying}\
	\eqref{eq:strength-target}\bigr\}\to0\label{eq:strength-exact}
\end{equation}
is exact as a sequence of sheaves, where $\Pot$ is the sheaf of quadruples
$(\mathbb V,\mathbb W,\bar{\mathbb V},\bar{\mathbb W})$ satisfying
$\mathbb V+\bar{\mathbb V}=\mathbb W+\bar{\mathbb W}$, $\G$ is the sheaf of
four semichiral gauge parameters acting by \eqref{eq:complex-gauge}, and $\str$ is the map
\eqref{eq:strength}. Every local chiral $f$ and twisted-chiral $\tilde f$
comes from SVM potentials, and if all strengths vanish the potentials are pure
gauge. The SVM action is quadratic in $f$ and $\tilde f$ \cite{NewVectors},
and although both come from the same three potentials, they satisfy no
local relation. The Bianchi identities
express the curvature components with a vector index as first and second
spinor derivatives of the strengths.

\paragraph{The SVM and the sum of the ordinary and twisted vectors.}
Locally the degree-zero cohomology of the SVM is that of an ordinary and a
twisted vector together \eqref{eq:mv}. This does not identify the full gauge
complexes: the direct sum has two lines of constant gauge transformations in
degree $-1$, whereas the SVM has one. This is compatible with the statement
that the SVM is not reducible to these two multiplets \cite{GatesMerrell},
which concerns the gauge structure: three real potentials rather than two,
semichiral parameters, and a single gauged symmetry.

\subsection{Reductions of the semichiral vector}\label{sec:reductions}
Imposing $f=0$ on the SVM should leave the ordinary vector multiplet, and
imposing $\tilde f=0$ the twisted one \cite{GatesMerrell}, \cite[Section~3.1]{SVM}.
In the language of Section~\ref{sec:gauge} this is immediate: $f$ and its
conjugate carry the curvature components $\lbp\lbm$ and $\lp\lm$, so imposing
$f=0$ removes them from the allowed set and leaves $\{\lp\lbm,\lbp\lm\}$, the
ordinary vector ring. In terms of partial potentials, $f=0$ says that
$\mathbb V=\Omega_{\bpl}-\Omega_{\bmi}$ is pure gauge: the two barred axis partial
potentials merge into one for the barred plane, likewise for the unbarred
ones, and two plane partial potentials are exactly the data of the ordinary
vector.

On the module side this is the surjection
$q:M_V\twoheadrightarrow M_{\mathbb V}$, which restricts functions on
two planes to the four axes they contain. On constrained superfields, a surjection of
rings gives an inclusion, because quotienting by more monomials leaves fewer
constraints: restriction $M_\Phi\to M_{\mathbb X_L}$ reconstructs the
inclusion of chiral into left semichiral superfields. We check that $q$ gives
the reduction on the full gauge complexes and not only on the rings. We
work with complexified complexes. Write $C_{\mathbb V}=[\G\to\Pot]$ for the
SVM gauge complex \eqref{eq:complex-gauge}, in which
$\bar{\mathbb V},\bar{\mathbb W},\bar{\mathbb X}_L,\bar{\mathbb X}_R$ are
independent and $\mathbb V+\bar{\mathbb V}=\mathbb W+\bar{\mathbb W}$; $C_f$
for its subcomplex with $f=\bar f=0$; and
$C_V=[\bar\Ch\oplus\Ch\to\U]$, $(\bar\Lambda',\Lambda)\mapsto\Lambda-\bar\Lambda'$,
for the ordinary vector, written for the potential $U=iV$, the negative of
the potential $-iV$ of Section~\ref{sec:cones}.

The claim is that the map $\iota_f:C_V\to C_f$ given by
\begin{equation}
	(\bar\Lambda',\Lambda)\mapsto(\mathbb X_L,\mathbb X_R,\bar{\mathbb X}_L,\bar{\mathbb X}_R)
	=(\Lambda,\Lambda,\bar\Lambda',\bar\Lambda'),\qquad
	U\mapsto(\mathbb V,\mathbb W,\bar{\mathbb V},\bar{\mathbb W})=(0,U,0,-U)
	\label{eq:reductions}
\end{equation}
is a local quasi-isomorphism, and that it is the reconstruction of the ring
surjection $q$. It is a chain map: $\delta\mathbb W=\mathbb X_L-\bar{\mathbb X}_R
	=\Lambda-\bar\Lambda'=\delta U$, $\delta\bar{\mathbb W}=-\delta U$ and
$\delta\mathbb V=\delta\bar{\mathbb V}=0$. Under it the remaining strength
$\tilde f=\bD_+D_-U$ is the twisted-chiral strength of the ordinary vector;
the partially conjugated statement holds with $f$ replaced by $\tilde f$ and the ordinary
vector by the twisted one. We check this in three steps.

\emph{(i) With $f=0$, the potential $\mathbb V$ is pure gauge.} The superfields
$\mathbb X_L^0=\mathbb V-\bar\theta^+\bD_+\mathbb V$ and $\mathbb X_R^0=-\bar\theta^+\bD_+\mathbb V$ satisfy
$\bD_+\mathbb X_L^0=0$, because $\bD_+^2=0$, and $\bD_-\mathbb X_R^0=-\bar\theta^+\bD_+\bD_-\mathbb V=0$,
because $f=0$; this uses only that $\bD_\pm$ are odd derivations with
$\bD_+\bar\theta^+=1$ and $\bD_-\bar\theta^+=0$. Moreover $\mathbb X_L^0-\mathbb X_R^0=\mathbb V$, so the
gauge transformation with parameters $(\mathbb X_L,\mathbb X_R)=(-\mathbb X_L^0,-\mathbb X_R^0)$ sets $\mathbb V=0$.
Conjugating the construction, $\bar f=0$ makes $\bar{\mathbb V}$ pure gauge
under $(\bar{\mathbb X}_L,\bar{\mathbb X}_R)$.

\emph{(ii) The quotient by the ordinary vector is acyclic.} With
$\mathbb V=\bar{\mathbb V}=0$ the relation
$\mathbb V+\bar{\mathbb V}=\mathbb W+\bar{\mathbb W}$ gives $\bar{\mathbb W}=-\mathbb W$. A
gauge transformation preserves $\mathbb V=0$ exactly when
$\mathbb X_L=\mathbb X_R$; such a parameter is both left and right semichiral,
hence chiral, $\mathbb X_L=\mathbb X_R=\Lambda$. Likewise $\bar{\mathbb V}=0$ is
preserved exactly when $\bar{\mathbb X}_L=\bar{\mathbb X}_R=\bar\Lambda'$ is
antichiral. These parameters act by $\delta\mathbb W=\Lambda-\bar\Lambda'$, so
$\iota_f$ is injective with image the subcomplex $\mathbb V=\bar{\mathbb V}=0$. The
quotient $C_f/\iota_f(C_V)$ is the two-term complex
\[
	\bigl[\G/\iota_f(\bar\Ch\oplus\Ch)
		\xrightarrow{\ (\mathbb X_L-\mathbb X_R,\ \bar{\mathbb X}_L-\bar{\mathbb X}_R)\ }
		\ker\bD_+\bD_-\oplus\ker D_+D_-\bigr],
\]
since modulo $\iota_f(\U)$ a potential in $C_f$ is determined by
$(\mathbb V,\bar{\mathbb V})$. Its differential is onto by (i), and injective
because $\mathbb X_L=\mathbb X_R$ and $\bar{\mathbb X}_L=\bar{\mathbb X}_R$ force the
parameter into $\iota_f(\bar\Ch\oplus\Ch)$. The quotient is therefore acyclic,
and $\iota_f$ is a quasi-isomorphism.

\emph{(iii) The module side.} The surjection $q$ restricts functions on the
two planes to the four axes they contain. On the restriction sequences
\eqref{eq:restriction} it is
\begin{equation}
	q_{\rm par}(p,p')=(p|_{\lp},\,p'|_{\lbp},\,p|_{\lm},\,p'|_{\lbm}),\qquad
	q_{\rm pot}(y,y')=(y,y',y,y'),\label{eq:reduction-square}
\end{equation}
for $p\in\C[\lp,\lm]$ and $p'\in\C[\lbp,\lbm]$, with the axes in the order
$(+,\bpl,-,\bmi)$; restriction commutes with evaluation at the origin, so
$q[1]$ is the induced map of cones. On parameters $q_{\rm par}$ says $\mathbb X_L=\mathbb X_R=\Lambda$
and $\bar{\mathbb X}_L=\bar{\mathbb X}_R=\bar\Lambda'$; on partial potentials $q_{\rm pot}$ says
$\Omega_+=\Omega_-=\Omega$ and $\Omega_{\bpl}=\Omega_{\bmi}=\bar\Omega$.
By \eqref{eq:svm-frames} this is $\mathbb V=\bar{\mathbb V}=0$ and
$\mathbb W=-\bar{\mathbb W}=\bar\Omega-\Omega=U$, the map $\iota_f$.

On the real form, $\bar\Lambda'=\bar\Lambda$ and $U=iV$ with $V$ real, and
\eqref{eq:reductions} becomes $\Lambda\mapsto(\mathbb X_L,\mathbb X_R)=(\Lambda,\Lambda)$,
$V\mapsto(\mathbb V,\mathbb W)=(0,iV)$, with $\delta V=i(\bar\Lambda-\Lambda)$ and
remaining strength $\tilde f=i\bD_+D_-V$.

The map $q$ and its partial conjugate are the two legs of the sequence \eqref{eq:mv}. A nonzero constant value of $f$ gives an affine
version of the same system, which appears as a twisted mass in
\cite[Section~3.2, (3.18)--(3.20)]{SVM}; it is not the reconstruction of a
ring map. Section~3 of \cite{SVM} also introduces the constrained semichiral
vector multiplet (CSVM); we do not identify it with a module here, and it is
not included in Table~\ref{tab:dictionary}.

\subsection{The large vector multiplet}\label{sec:lvm}
The large vector multiplet has chiral and twisted-chiral gauge parameters
\cite{NewVectors}. As in Section~\ref{sec:cones}, we build a gauge
multiplet from the rings whose reconstructions are its gauge parameters
(Section~\ref{sec:matter}). Here these are the coordinate rings of the
chiral plane, the twisted-chiral plane and their conjugates,
\[
	\C[Z_c]=\C[\lbp,\lbm],\quad \C[Z_{\bar c}]=\C[\lp,\lm],\quad
	\C[Z_t]=\C[\lbp,\lm],\quad \C[Z_{\bar t}]=\C[\lp,\lbm],
\]
which reconstruct the chiral, antichiral, twisted-chiral and
twisted-antichiral superfields. These four planes are the irreducible
components of the nilpotence variety $Y$.

Unlike the pieces of the other vector multiplets, these planes meet along
coordinate axes and not only at the origin. Each ghost lies on two of them:
$\lp$, for instance, lies on both $\C[Z_{\bar c}]$ and $\C[Z_{\bar t}]$. Independent
plane connections need not agree along the shared axes, so they do not in
general define a single connection. The argument leading to
\eqref{eq:frames} therefore does not apply directly. Instead, \emph{each plane
	carries its own flat connection}, $\Gamma_{i,a}=D_i\Omega_a$ for
$\lambda_i$ on $\C[Z_a]$, and on each shared axis $\lambda_i$ the two planes
through it give two components $\Gamma_{i,a}$ that need not agree. The large
vector multiplet is these four flat connections modulo a common gauge
transformation, and its gauge-invariant content is their mismatch along the
four shared axes (Theorem~\ref{thm:lvm-strengths}). A single connection on
all four planes would have no mismatch and would be flat
(see \eqref{eq:flatness}). The plane connections on $Z_c$ and $Z_{\bar c}$
together have one component along each ghost and form a connection
$\nabla$; those on $Z_t$ and $Z_{\bar t}$ form a second connection
$\widetilde\nabla$. These are the two sets of covariant derivatives of the LVM
\cite{Nonabelian}, acting on chiral and on twisted-chiral matter
respectively. They share one gauge parameter $\kappa$, which shifts all four
partial potentials, reflecting the single gauged symmetry. With independent
parameters for $\nabla$ and $\widetilde\nabla$, only $\Omega_c-\Omega_{\bar c}$
and $\Omega_t-\Omega_{\bar t}$ would be gauge invariant, giving an ordinary and
a twisted vector multiplet; the common $\kappa$ produces the third
potential and hence the LVM.

We use the same evaluation-at-the-origin construction as in
Section~\ref{sec:cones}, but match the plane functions only at the origin,
not along their shared axes.
The partial potentials $\Omega_a$ are fixed up to $\eta_c=\Lambda$
chiral, $\eta_t=\tilde\Lambda$ twisted chiral and their conjugates.
Modulo a common shift, the four of them have three independent differences,
\[
	\mathcal V=\Omega_c-\Omega_t,\qquad \mathcal W=\Omega_c-\Omega_{\bar t},\qquad
	\bar{\mathcal V}=\Omega_{\bar c}-\Omega_{\bar t},\qquad
	\bar{\mathcal W}=\Omega_{\bar c}-\Omega_t,
\]
again with $\mathcal V+\bar{\mathcal V}=\mathcal W+\bar{\mathcal W}$ and with
barred fields independent until the reality condition is imposed. On the
real form
\begin{equation}
	\delta \mathcal V=\Lambda-\tilde\Lambda,\qquad\delta \mathcal W=\Lambda-\bar{\tilde\Lambda}.\label{eq:lvm-gauge}
\end{equation}
On the module side, with $N=\C[Z_c]\oplus \C[Z_{\bar c}]\oplus \C[Z_t]\oplus \C[Z_{\bar t}]$,
the restriction sequence is
\begin{equation}
	0\to M_{\mathcal V}\to N\xrightarrow{\ \mathrm{ev}\ }k^4/k\to0,
	\label{eq:plane-sequence}
\end{equation}
where
\[
	M_{\mathcal V}=\C[Z_c]\times_k\C[Z_{\bar c}]\times_k\C[Z_t]\times_k\C[Z_{\bar t}]
	=\bigl\{(p_a):p_a(0)\ \text{all equal}\bigr\}.
\]
With the componentwise product, $M_{\mathcal V}$ is a commutative $\OY$-algebra.
Restriction to the planes embeds $\OY$ into it, and it lies inside $N$,
the normalization of $\OY$:
\[
	\OY\subset M_{\mathcal V}\subset N.
\]
Unlike $\OY$, the algebra $M_{\mathcal V}$ matches the plane functions only at the
origin and not along the four axes where the planes meet. For the semichiral
vector the pieces meet only at the origin, and the fiber product is the
coordinate ring $M_{\mathbb V}$; for the large vector it is not a quotient of
$\OY$ at all. Its minimal graded free resolution over $R$ is
\[
	0\to R(-4)^3\to R(-3)^{12}\to R(-2)^{14}
	\to R\oplus R(-1)^4\to M_{\mathcal V}\to0.
\]
The Betti numbers follow from Lemma~\ref{lem:betti} in
Appendix~\ref{app:betti}. In particular, the module is not cyclic: it has
one minimal generator in weight zero and four in weight one. We may take
$\mathbf1=\sum_ae_a$, $\lp e_{\bar c}$, $\lm e_{\bar c}$,
$\lbp e_c$, $\lbm e_c$, where $e_a$ is the unit of $\C[Z_a]$.
For each ghost the generator supported on the other plane containing it
is $\lambda_i\mathbf1$ minus the listed generator. This gives all eight
weight-one monomials; multiplication within each plane gives the rest.

To identify the reconstructed multiplet, apply the cone argument of
Theorem~\ref{thm:vectors} to \eqref{eq:plane-sequence}. This gives the
LVM gauge complex
\begin{equation}
	\A(M_{\mathcal V})[1]\simeq[\G_{\mathcal V}\to\U^3]_{-1,0},
	\label{eq:lvm-complex}
\end{equation}
with the gauge transformations \eqref{eq:lvm-gauge} and their barred
counterparts, obtained from the displayed differences of partial potentials.
Here $\G_{\mathcal V}$ is the sheaf of quadruples
$(\Lambda,\bar\Lambda,\tilde\Lambda,\bar{\tilde\Lambda})$, respectively chiral,
antichiral, twisted chiral and twisted antichiral. On the real form only
$\Lambda,\tilde\Lambda$ are independent. Indeed, the sequence is exact by
definition of $M_{\mathcal V}$, $\mathrm{ev}$ is onto, and \eqref{eq:linsupp}
identifies the parameters. A parameter with zero variation has
$\Lambda=\tilde\Lambda=\bar{\tilde\Lambda}=\bar\Lambda$; this common parameter
is annihilated by every spinor derivative, so the constant gauge transformations form the locally
constant line in degree $-1$, real on the real form.

The gauge-invariant content of the LVM follows from a second exact sequence,
which compares $M_{\mathcal V}$ with $\OY$. Each ghost $\lambda_i$ spans the
common axis of exactly two planes, one chiral or antichiral, $a(i)\in\{c,\bar c\}$,
and one twisted, $b(i)\in\{t,\bar t\}$. Let
$\rho_i(p)=p_{a(i)}|_{\lambda_i}-p_{b(i)}|_{\lambda_i}$ be the difference
of the two plane functions on their common axis.

\begin{theorem}\label{thm:lvm-strengths}
	The sequence
	\begin{equation}
		0\to \OY\to M_{\mathcal V}\xrightarrow{\ \oplus_i\rho_i\ }
		\bigoplus_i\lambda_i\C[\lambda_i]\to0\label{eq:lvm-filtration}
	\end{equation}
	is exact. Let $\Pot$ be the sheaf of quadruples
	$(\mathcal V,\mathcal W,\bar{\mathcal V},\bar{\mathcal W})$ satisfying
	$\mathcal V+\bar{\mathcal V}=\mathcal W+\bar{\mathcal W}$. Consequently,
	\begin{gather*}
		0\to\C\to\G_{\mathcal V}\to\Pot\xrightarrow{\ \str\ }
		\bigoplus_i\Pi\ker D_i\to0,\\
		\str(\mathcal V,\mathcal W,\bar{\mathcal V},\bar{\mathcal W})=\bigl(D_+\bar{\mathcal V},\ \bD_+\mathcal V,\ D_-\bar{\mathcal W},\ \bD_-\mathcal W\bigr),
	\end{gather*}
	is locally exact as a sequence of sheaves: LVM potentials modulo gauge are
	exactly four odd superfields $\mathbb G_i$ with $D_i\mathbb G_i=0$, subject
	to no further local differential relation in the complexified setting.
	On the real form the strengths are paired by conjugation.
\end{theorem}
\begin{proof}
	\emph{Exactness.} A tuple in $M_{\mathcal V}$ has equal values at the origin, so
	each $\rho_i$ has no constant term. The pure power $\lambda_i^we_a$, $w\ge1$,
	with $e_a$ the unit of $\C[Z_a]$ and $a=a(i)$, lies in $M_{\mathcal V}$, restricts to
	zero on the other axis of $\C[Z_a]$, and has $\rho_i=\lambda_i^w$ and
	$\rho_j=0$ for $j\neq i$; so $\oplus_i\rho_i$ is onto. Since $I_T$ is
	radical, $\OY$ is the ring of functions on the union of the four planes and
	injects into $M_{\mathcal V}$; its image lies in the kernel, because
	restrictions of one function agree on the axes. Equality follows by counting
	dimensions weight by weight. In weight zero the three terms are $k$, $k$
	(the diagonal constants) and $0$. In weight $w\ge1$, $\OY$ is spanned by the
	$4w$ monomials divisible by neither $\lp\lbp$ nor $\lm\lbm$, $M_{\mathcal V}$
	has dimension $4(w+1)$, four planes with $w+1$ monomials each, and the
	target has dimension $4$. Since $\oplus_i\rho_i$ is onto, its kernel has
	dimension $4w$ and equals $\OY$.

	\emph{Reconstruction.} Reconstructing the shifted sequence
	\eqref{eq:lvm-filtration} gives a long exact sequence in cohomology.
	Locally, and using \eqref{eq:axis} for the second term,
	\[
		H^\bullet\bigl(\A(\OY)[1]\bigr)=\C\ \text{in degree }-1,\qquad
		H^\bullet\bigl(\A(\lambda_i\C[\lambda_i])[1]\bigr)=\Pi\ker D_i\
		\text{in degree }0.
	\]
	Hence
	\[
		H^{-1}\bigl(\A(M_{\mathcal V})[1]\bigr)=\C,\qquad
		H^{0}\bigl(\A(M_{\mathcal V})[1]\bigr)\cong\bigoplus_i\Pi\ker D_i .
	\]
	It remains to identify this isomorphism with $\str$. Since no plane contains
	a torsion pair, a degree-zero cocycle of $\A(M_{\mathcal V})[1]$ is a family of
	flat connections on the four planes,
	\[
		\Gamma_{j,a}=D_j\Omega_a\qquad\text{for }\lambda_j\text{ on }\C[Z_a] .
	\]
	Each axis $\lambda_i$ lies on two of the planes, $\C[Z_{a(i)}]$ and $\C[Z_{b(i)}]$,
	and $\rho_i$ takes the difference of the two $\lambda_i$-components:
	\[
		\Gamma_{i,a(i)}-\Gamma_{i,b(i)}=D_i\bigl(\Omega_{a(i)}-\Omega_{b(i)}\bigr).
	\]
	By the definitions of $\mathcal V,\mathcal W,\bar{\mathcal V},\bar{\mathcal W}$,
	this is $\str$.
\end{proof}

The LVM strengths are thus differences of the two connections along a shared
direction, objects of weight one, whereas the SVM strengths are curvatures of
a single connection, of weight two. On the real form the four strengths pair
into two complex odd semichiral superfields, $\bD_+\mathcal V$ and $\bD_-\mathcal W$, and their
conjugates, the semichiral spinor field strengths of
\cite{NewVectors,Nonabelian}. The
filtration \eqref{eq:lvm-filtration} separates the gauge skeleton $\OY$, whose
reconstruction carries only the constant gauge transformations, from the four axis modules, whose
reconstructions carry the strengths.

The resulting $16+16$ real component count can also be read from the
resolution above. After the shift by $[1]$, the weight-zero generator gives
the scalar gauge parameter in degree $-1$. The other classes have ghost
degree zero, and the parity shift gives a class of Koszul degree $i$
parity $i+1$ modulo two. Thus $\Tor_1$ and $\Tor_3$ give $14+3$ bosonic
components, while the four weight-one classes of $\Tor_0$ together with
$\Tor_2$ give $4+12$ fermionic components. Removing one bosonic gauge
direction at generic momentum gives $16+16$. This agrees with subtracting
the $8+8$ components of the chiral and twisted-chiral gauge parameters
from the $24+24$ of the three real unconstrained potentials. These
subtractions concern the gauge image at generic momentum; the full gauge
complex retains the constant gauge transformations in degree $-1$.

\paragraph{Constrained large vector multiplets.}\label{sec:clvm}
Agreement along only some of the shared axes already occurs in the
literature. Bykov, Lindstr\"om and Ro\v cek
\cite[Section~3, (3.6)]{ConstrainedLVM} identify the gauge multiplet of
\cite{BKK} as an LVM constrained by $\bD_+V_L=0$ and its conjugate.
Their $V_L$ is our $\mathcal V$, so this sets the two strengths associated
with the $+$ and $\bpl$ axes to zero. In our construction, this means
regluing the plane functions along precisely these two axes. The remaining
strength is the complex odd semichiral superfield $\bD_-\mathcal W$, with
$8+8$ real components. The interpretation as a partial dualization is
established in \cite{ConstrainedLVM}; here we identify the corresponding
module and its reconstructed gauge complex.

More generally, setting a strength to zero enforces agreement along its
axis. For $S\subseteq\{+,\bpl,-,\bmi\}$ let
\[
	M_S=\ker\Bigl(\bigoplus_{i\in S}\rho_i\colon M_{\mathcal V}\to\bigoplus_{i\in S}\lambda_i\C[\lambda_i]\Bigr),
\]
the functions on the four planes that agree along the axes in $S$ (and at
the origin). Agreement is preserved by products, so $M_S$ is an
$\OY$-algebra with $\OY=M_{\{+,\bpl,-,\bmi\}}\subseteq M_S\subseteq M_\emptyset=M_{\mathcal V}$.
Geometrically, $M_S$ is obtained from the normalization by regluing only the axes in $S$.

\begin{proposition}\label{prop:clvm}
	\begin{enumerate}[label=(\roman*)]
		\item The sequences $0\to M_S\to M_{\mathcal V}\to\bigoplus_{i\in S}\lambda_i\C[\lambda_i]\to0$ and
		      $0\to\OY\to M_S\to\bigoplus_{i\notin S}\lambda_i\C[\lambda_i]\to0$ are exact.
		\item Locally, $\A(M_S)[1]\simeq[\G_{\mathcal V}\to\Pot_S]_{-1,0}$, where $\Pot_S$
		      consists of the LVM potentials with $\str_i=0$ for $i\in S$. The cohomology is
		      $\C$ in degree $-1$ and $\bigoplus_{i\notin S}\Pi\ker D_i$ in degree $0$.
	\end{enumerate}
\end{proposition}
\begin{proof}
	(i) Surjectivity uses the elements $\lambda_i^we_{a(i)}$ from the proof of
	Theorem~\ref{thm:lvm-strengths}; exactness in the middle follows from the
	definition and from \eqref{eq:lvm-filtration}.
	(ii) Reconstruct the first sequence, shifted by $[1]$. Its target has
	cohomology only in degree $0$, where the induced map is the projection of
	$\str$ onto the components in $S$. By Theorem~\ref{thm:lvm-strengths},
	this projection is locally onto. Its fiber is therefore represented by
	the gauge complex with potentials in $\Pot_S$, and the long exact sequence
	gives the stated cohomology.
\end{proof}

Conjugation exchanges $\lambda_\pm\leftrightarrow\bar\lambda_\pm$, so the
choices compatible with the real form are $S=\emptyset$ (the LVM),
$\{+,\bpl\}$ (the example above), $\{-,\bmi\}$ (the same with $+\leftrightarrow-$), and all four
axes (flat). Nonreal choices give complex multiplets, with one odd
semichiral strength for each axis not in $S$; we do not know whether they
have a physical role.

\paragraph{Unglued multiplets with semichiral parameters.}
The fiber-product construction also applies when the matter supports are
nested. Consider chiral and semichiral matter of both chiralities,
together with their conjugates, and let every kind of matter keep its own
gauge parameter. The pieces are the planes $Z_c,Z_{\bar c}$ and the four
axes, each axis lying in the plane $Z_{a(i)}$ of the same chirality type,
and the gauge module is the fiber product
$M_{V\mathbb V}=M_V\times_kM_{\mathbb V}$. Gluing the axes into their planes
instead recovers $M_V$, the ordinary coupling of semichiral matter with
chiral parameters. Let $\rho_i(p)=p_i-p_{a(i)}|_{\lambda_i}$ compare an axis
function with the restriction of the plane containing it. Counting
dimensions weight by weight, as in the proof of
Theorem~\ref{thm:lvm-strengths}, shows that
$0\to M_V\to M_{V\mathbb V}\xrightarrow{\oplus_i\rho_i}
\bigoplus_i\lambda_i\C[\lambda_i]\to0$ is exact. Since $\A(M_V)[1]$ has no
cohomology in degree one (Proposition~\ref{prop:strengths}), locally
$H^{-1}(\A(M_{V\mathbb V})[1])=\C$, and $H^0$ is an extension of
$\bigoplus_i\Pi\ker D_i$, spanned by the odd strengths
$\mathbb H_i=D_i(\Omega_i-\Omega_{a(i)})$, by $H^0(\A(M_V)[1])$, spanned by
the strengths $\sigma,\bar\sigma$ of the ordinary vector. The curvatures of the axis
connection are descendants; for instance
$f=-(\bD_-\mathbb H_{\bpl}+\bD_+\mathbb H_{\bmi})$. The multiplet has five
real potentials and $20+20$ real components at generic momentum, which is
$4+4$ from $\sigma$ and $16+16$ from the $\mathbb H_i$.
The free resolution is
\begin{equation}
	0\to R(-4)^5\to R(-3)^{16}\to R(-2)^{16}\to R\oplus R(-1)^4\to M_{V\mathbb V}\to0.
	\label{eq:res-VSV}
\end{equation}
The maximal choice
takes all four planes, chiral and twisted chiral, together with all four
axes. Its module is $M_{\max}=M_{\mathcal V}\times_kM_{\mathbb V}$, and the
reduced union of the supports is $\OY$. Each axis now lies on two planes
and is a piece in its own right, so the same count gives
$0\to\OY\to M_{\max}\to\bigoplus_i(\lambda_i\C[\lambda_i])^{2}\to0$.
Because $\A(\OY)[1]$ contributes only the constants, locally
\[
	H^0\bigl(\A(M_{\max})[1]\bigr)\cong\bigoplus_i(\Pi\ker D_i)^{2},
\]
spanned by the strengths $\mathbb G_i$ of the large vector multiplet and
by the $\mathbb H_i$. This gives eight odd semichiral strengths, four
complex ones on the real form, seven real potentials, and $32+32$ real
components with free resolution 
\begin{equation}
	0\to R(-4)^7\to R(-3)^{24}\to R(-2)^{26}\to R\oplus R(-1)^8\to M_{\max}\to0.
	\label{eq:res-max}
\end{equation}
Both multiplets are real, and their gauge-invariant content
follows from the module without further input. We have not constructed
superspace presentations of these multiplets, and their relevance for
generalized K\"ahler sigma models remains open.

\paragraph{Algebra structures and connection geometry.}
The gauge modules considered here are commutative algebras over the CE
algebra. Their unshifted reconstructions $\A(M)$ therefore carry products
compatible with the differential. This additional structure connects our
presentations to the multiplicative resolutions and Koszul-dual description
of canonical multiplets developed in
\cite[Sections~3.1--3.5]{Canonical}.

The distribution-based formulation of Hahner and Saberi
\cite{HahnerSaberi} provides the geometric setting: a distribution on a
supermanifold induces a filtration of its de Rham complex. In flat
superspace, the Levi form of the odd distribution is the torsion map
defining the pure spinor quadrics. Our constructions keep this distribution
fixed and replace its derivatives by abelian covariant derivatives.
The quotient rings encode which curvature components must vanish; the
resulting connections are flat on each linear piece of the support, rather
than necessarily flat on all of superspace. The partial potentials of
Section~\ref{sec:cones} are local trivializations along those directions, and their
differences recover the superspace gauge potentials. For the LVM, the
fiber-product algebra instead records separate partial connections and
their mismatches along shared axes.

\section{A dual presentation of complex-linear matter}\label{sec:duality}
The complex-linear multiplet of Section~\ref{sec:linear} was constructed from
a degree-two Ext class. Here we give a second presentation: up to shift,
the same object is the Gorenstein dual of the augmentation ideal of the
chiral module. The superspace motivation is the annihilator pairing that
enters physical chiral--linear duality. We use this pairing only as
motivation and do not assume that reconstruction intertwines the two notions
of duality in general.

\paragraph{The superspace pairing.}
In the familiar chiral--complex-linear duality
\cite{GatesSiegel,DeoGates,GGW}, a chiral Lagrange multiplier $\Phi$
imposes the complex-linear constraint through the pairing
$\int d^2x\,d^4\theta\,U\Phi=\pm\int d^2x\,d^2\theta\,(\bD_+\bD_-U)\Phi$.%
\footnote{For the scalar--tensor analog with real linear superfields, see
	\cite{LR83}. The duality relates the free theories, not the $2+2$ and
	$6+6$ off-shell multiplets; this distinction is emphasized in
	\cite{Hubsch}.}
With compactly supported chiral test superfields, nondegeneracy of this
pairing identifies the annihilator of $\Ch$ in $\U$ with $\ker(\bD_+\bD_-)$,
which is therefore naturally paired with the quotient $\U/\Ch$.

\paragraph{The dual of the augmentation ideal.}
On the module side, consider
\[
	0\to\mathfrak j\to M_\Phi\to k\to0,
	\qquad \mathfrak j=(\lbp,\lbm)\subset M_\Phi.
\]
Reconstruction turns $M_\Phi\to k$ into the inclusion of chiral into
unconstrained superfields, so $\A(\mathfrak j)\simeq(\U/\Ch)[-1]$.
This motivates applying the contravariant derived dual
$\mathbb D=\RHom_\OY(-,\OY)$ to $\mathfrak j$, rather than to $M_\Phi$
itself.

\begin{theorem}\label{thm:dual}
	Let $\mathbb D=\RHom_\OY(-,\OY)$. Then
	\[
		\mathbb D(M_\Phi)\simeq(\lbp\lbm)\simeq M_\Phi(-2),\qquad
		\mathbb D(k)\simeq k[-2],\qquad
		\mathbb D(\mathfrak j)\simeq M_\Sigma[-1],
	\]
	with $M_\Sigma$ the complex-linear module \eqref{eq:MSigma}.
	Consequently, applying $\A$ to the bounded model $M_\Sigma[-1]$,
	$\A(\mathbb D(\mathfrak j))[1]\simeq\ker(\bD_+\bD_-:\U\to\Ch)$.
\end{theorem}
\begin{proof}
	Write $\OY=B_+\otimes_\C B_-$ with $B_+=\C[\lp,\lbp]/(\lp\lbp)$ and
	$B_-=\C[\lm,\lbm]/(\lm\lbm)$. The module $B_+/(\lp)$ has the periodic
	resolution with maps alternating between $\lp$ and $\lbp$, and since
	$\Ann(\lp)=(\lbp)$ and $\Ann(\lbp)=(\lp)$, its dual is exact except in degree
	zero, where it is $(\lbp)$. Tensoring with the analogous statement for $B_-$
	gives $\mathbb D(M_\Phi)=(\lbp\lbm)\cong M_\Phi(-2)$. For $k$, the two relations form a
	regular sequence, and change of rings from $R$, where $\Ext^4_R(k,R)=k(4)$,
	gives $\Ext^\bullet_\OY(k,\OY)=k$ in degree two and weight zero.
	The complete-intersection formula for the canonical module gives internal
	shift $2+2-4=0$: there are two quadratic relations and four weight-one
	variables. The equivariant character cancels for the same reason.
	If $\mathrm{ch}_i$ denotes the character of $\lambda_i$, the determinant character
	of the relation space is
	$(\mathrm{ch}_+\mathrm{ch}_{\bpl})(\mathrm{ch}_-\mathrm{ch}_{\bmi})$, equal to
	$\mathrm{ch}_+\mathrm{ch}_{\bpl}\mathrm{ch}_-\mathrm{ch}_{\bmi}$ for the variable space.
	Their ratio is trivial, so no Lorentz or R-character twist remains.
	Thus $\OY$ is Gorenstein with canonical module $\OY$, without an internal
	shift or character twist, and for finitely generated graded $\OY$-modules the natural map
	$M\to\mathbb D\mathbb D(M)$ is a quasi-isomorphism \cite{BrunsHerzog}.

	Dualizing $\mathfrak j\to M_\Phi\to k$ gives a triangle
	$k[-2]\to M_\Phi(-2)\to\mathbb D(\mathfrak j)\to k[-1]$. Its first arrow is a
	class in $\Ext^2_\OY(k,M_\Phi)_{-2}$, so by \eqref{eq:ext} it is a
	multiple $ac$, $a\in\C$, of the complex-linear class. It is nonzero: otherwise
	$\mathbb D(\mathfrak j)\simeq M_\Phi(-2)\oplus k[-1]$ would split, and by
	biduality $\mathfrak j\simeq\mathbb D\mathbb D(\mathfrak j)$ would have
	cohomology in two degrees, whereas $\mathfrak j$ is a module. Hence
	$\mathbb D(\mathfrak j)$ is the cone of a nonzero multiple of $c[-2]$, that
	is $\Cone(c)[-2]=M_\Sigma[-1]$.
\end{proof}

It is the augmentation ideal, not the chiral module itself,
whose dual yields complex-linear matter: $\mathbb D(M_\Phi)\simeq M_\Phi(-2)$
is again chiral. Both statements match the superspace pairings at the level
of objects. Chiral superfields pair with chiral superfields through chiral
superspace, $\int d^2x\,d^2\theta\,\Phi\Phi'$, so the chiral multiplet is its
own partner, as $\mathbb D(M_\Phi)\simeq M_\Phi(-2)$ is; the shift $(-2)$ has the
weight of the chiral measure. The full-superspace pairing makes
$\ker(\bD_+\bD_-)$ the partner of $\U/\Ch=\A(\mathfrak j)[1]$, as
$\mathbb D(\mathfrak j)\simeq M_\Sigma[-1]$ does. Whether $\A$ intertwines
$\mathbb D$ with the duality of multiplets of \cite[Corollary~4.9]{Derived}
and \cite{Canonical} for these modules we leave open. The result is specifically two-dimensional: the
four-dimensional ring $\OYd{4}$ is not Gorenstein, so the argument does not
apply unchanged.\footnote{The ring $\OYd{4}$ is the union of two
	planes meeting at the origin. Its punctured spectrum is disconnected,
	so $\operatorname{depth}\OYd{4}=1<2=\dim\OYd{4}$; hence it is not
	Cohen--Macaulay \cite[Section~4.4]{Perspectives}. A possible replacement
	using a dualizing complex or the full CE algebra is not developed here.}

\section{Conclusions}\label{sec:conclusions}
We have given a dictionary for matter and gauge multiplets, including several
less familiar ones, and shown how to realize them in the pure spinor
formalism. In particular, constraints, potentials and gauge transformations
acquire a clear interpretation at the level of the modules corresponding to
the multiplets. With the same techniques we have also proposed new
multiplets, which we leave for future study. Table~\ref{tab:dictionary}
collects the correspondence, the dictionary \eqref{eq:dictionary} of the
introduction.

\begin{table}[H]
	\centering\small
	\begin{tabular}{@{}llll@{}}\toprule
		multiplet                                                      & module                                            & superspace presentation                                                                       & where                  \\\midrule
		unconstrained scalar                                           & $k$                                               & no constraint                                                                                 & \S\ref{sec:chiral}     \\
		chiral / twisted-chiral                                        & $M_\Phi$ / $M_\chi$                               & $\bD_\pm\Phi=0$ / $\bD_+\chi=D_-\chi=0$                                                       & \S\ref{sec:chiral}     \\
		semichiral $\mathbb X_L$ / $\mathbb X_R$                       & $M_{\mathbb X_L}$ / $M_{\mathbb X_R}$             & $\bD_+\mathbb X_L=0$ / $\bD_-\mathbb X_R=0$                                                   & \S\ref{sec:semichiral} \\
		conjugate semichiral $\bar{\mathbb X}_L$ / $\bar{\mathbb X}_R$ & $M_{\bar{\mathbb X}_L}$ / $M_{\bar{\mathbb X}_R}$ & $D_+\bar{\mathbb X}_L=0$ / $D_-\bar{\mathbb X}_R=0$                                           & \S\ref{sec:semichiral} \\
		complex-linear                                                 & $M_\Sigma=\Cone(c)[-1]$                           & $\bD_+\bD_-\Sigma=0$                                                                          & \S\ref{sec:linear}     \\
		$3d$ / $4d$ vector                                             & $\OYd{3}[1]$ / $\OYd{4}[1]$                       & $\delta\Gamma_\alpha=D_\alpha\kappa$ / $\delta V=i(\bar\Lambda-\Lambda)$                      & \S\ref{sec:cones}      \\
		$2d$ vector / twisted                                          & $M_V[1]$ / $M_{\widetilde V}[1]$                  & $\delta V=i(\bar\Lambda-\Lambda)$ / $\delta\widetilde V=i(\bar{\tilde\Lambda}-\tilde\Lambda)$ & \S\ref{sec:ordinary}   \\
		semichiral vector                                              & $M_{\mathbb V}[1]$                                & $\delta \mathbb V=\mathbb X_L-\mathbb X_R$, $\delta \mathbb W=\mathbb X_L-\bar{\mathbb X}_R$  & \S\ref{sec:svm}        \\
		large vector                                                   & $M_{\mathcal V}[1]$                               & $\delta \mathcal V=\Lambda-\tilde\Lambda$, $\delta \mathcal W=\Lambda-\bar{\tilde\Lambda}$    & \S\ref{sec:lvm}        \\\bottomrule
	\end{tabular}
	\caption{The dictionary between modules and superspace presentations. Here $M_\Phi=\C[\lbp,\lbm]$ is the chiral module and $c:k\to M_\Phi(-2)[2]$ is the Yoneda class of Section~\ref{sec:linear}. Throughout, $\Lambda$ denotes a chiral and $\tilde\Lambda$ a twisted-chiral gauge parameter.}
	\label{tab:dictionary}
\end{table}

Read from the module side, the dictionary has some features that are not
visible in superspace. The order of a constraint is the degree of a derived
morphism: second-order constraints such as $\bD_+\bD_-\Sigma=0$ do not come
from module maps, only from Ext classes, and the auxiliary fields of their
first-order formulations are the intermediate modules of a Yoneda product.
A gauge multiplet is determined by the matter it gauges: the geometry of the
matter supports fixes the allowed curvature, the potentials and the gauge
parameters at once. Where a single connection suffices, the gauge module is a
quotient of $\OY$; the large vector multiplet shows that this is not always
the case, and its module, which matches the plane functions only at the
origin, is not a quotient of $\OY$ at all. The same construction applied to
other matter collections produces the constrained large vector multiplets and
the unglued multiplets $M_{V\mathbb V}$ and $M_{\max}$, whose gauge-invariant
content follows from the module before any superspace presentation is known.

Our setting has two obvious restrictions, which leave open questions for
future work: all gauge multiplets are abelian, and supersymmetry is rigid
throughout, so none of the examples involves local supersymmetry or
supergravity. Further directions include:
\begin{itemize}
	\item finding superspace presentations of the unglued multiplets
	      $M_{V\mathbb V}$ and $M_{\max}$ of Section~\ref{sec:lvm}, testing the
	      fiber-product principle on further matter collections whose gauge
	      multiplets are not known, such as chiral with only left semichiral
	      matter, and placing the CSVM of \cite{SVM} in the dictionary;
	\item three dimensions with $\mathcal N=2$, where the non-torsion mixed
	      quadric $\mu=\varepsilon_{\alpha\beta}\lambda^\alpha\bar\lambda^\beta$ spans an
	      embedded point of the pure spinor scheme. Chiral matter gives the ring
	      $R/(\lambda^\alpha\bar\lambda^\beta)$, and the construction of
	      Theorem~\ref{thm:vectors} suggests the analogous presentation
	      $[\bar\Ch\oplus\Ch\to\U]$ for the $3d$ $\mathcal N=2$ vector; we expect $\mu$ to carry
	      its real linear field strength, which would need the full CE algebra. The
	      number of non-torsion mixed quadrics, $0$, $1$ and $2$ for $4d$
	      $\mathcal N=1$, $3d$ $\mathcal N=2$ and $2d$ $(2,2)$, equals the number of
	      real scalars in the vector multiplet;
	\item $(0,2)$ supersymmetry, where the translation ghost $\vmm$ is closed, so
	      that $\CE{(0,2)}$ is not quasi-isomorphic to its pure spinor ring. Chiral and
	      Fermi multiplets come from modules of linear support, and a linear $E$-term is the cone
	      of multiplication by a ghost; a module for the $(0,2)$ vector, which must
	      involve $\vmm$, and the restriction from $(2,2)$ by derived base change
	      remain open;
	\item nonabelian extensions, starting from the dg Lie algebra
	      $\A(M)\otimes\mathfrak g$ supplied by the commutative gauge algebra,
	      and comparison of its Maurer--Cartan equation with nonlinear
	      superspace gauge presentations; global extensions require
	      compatible gluing data.
\end{itemize}

\appendix
\section{The complex-linear Ext class and a finite CE model}\label{app:linear}

\subsection{The resolution of \texorpdfstring{$k$}{k}}\label{app:res-k}
The uniqueness calculation uses the minimal free resolution of $k$ over
$\OY$. Since $\OY=B_+\otimes_\C B_-$ with
$B_\pm=\C[\lambda_\pm,\bar\lambda_\pm]/(\lambda_\pm\bar\lambda_\pm)$, it is
the tensor product of the periodic resolutions
\[
	\cdots\to B_+(-2)^2
	\xrightarrow{\operatorname{diag}(\lbp,\lp)}B_+(-1)^2
	\xrightarrow{(\lp\ \lbp)}B_+\to\C\to0
\]
of the two factors, exact because $\Ann(\lp)=(\lbp)$ and $\Ann(\lbp)=(\lp)$.
Its ranks are $(1+2t+2t^2+\cdots)^2=1+4t+8t^2+12t^3+\cdots$:
\begin{equation}
	k\longleftarrow\OY\xleftarrow{\ d_1\ }\OY(-1)^4\xleftarrow{\ d_2\ }\OY(-2)^8
	\xleftarrow{\ d_3\ }\OY(-3)^{12}\longleftarrow\cdots.\label{eq:res-k}
\end{equation}
The generators $\xi_i$ of degree $-1$ map to $\lambda_i$. The syzygy module
$\Omega^2=\ker d_1$ is generated by the six Koszul syzygies
$K_{ij}=\lambda_i\xi_j-\lambda_j\xi_i$ and two new ones,
$u_{\pmpm}=\bar\lambda_\pm\xi_\pm$: $d_1u_{\pplus}=\lbp\lp$ vanishes in $\OY$
but not in $R$, so they are the module-side trace of the translation
ghosts. In the order
$(K_{+\bpl},K_{+-},K_{+\bmi},K_{\bpl-},K_{\bpl\bmi},K_{-\bmi},u_{\pplus},u_{\mminus})$,
with rows of $d_2$ indexed by $(\xi_+,\xi_{\bpl},\xi_-,\xi_{\bmi})$,
\begin{equation}
	d_2=\begin{pmatrix}
		-\lbp & -\lm & -\lbm & 0    & 0     & 0     & \lbp & 0    \\
		\lp   & 0    & 0     & -\lm & -\lbm & 0     & 0    & 0    \\
		0     & \lp  & 0     & \lbp & 0     & -\lbm & 0    & \lbm \\
		0     & 0    & \lp   & 0    & \lbp  & \lm   & 0    & 0
	\end{pmatrix}.
\end{equation}
By minimality, twelve relations generate the relations among these eight
generators. We need one property of them: in every relation, $K_{\bpl\bmi}$
is multiplied by an element of $(\lp,\lm)$. Indeed, the ideal of $\OY$ is
generated by monomials, so the resolution is graded by $\mathbb Z^4$, one
degree for each ghost. A relation in which $K_{\bpl\bmi}$ had a coefficient
in the barred ghosts alone would live in a multidegree involving only
$\lbp,\lbm$. In such multidegrees the resolution is that of
$\C[\lbp,\lbm]$, whose $\Tor_3$ vanishes: $\C[\lbp,\lbm]$ is an algebra
retract of $\OY$, and in these multidegrees the bar construction of $\OY$
only involves elements of $\C[\lbp,\lbm]$.

\subsection{The complex-linear class}\label{app:linear-class}
The class $c$ of Section~\ref{sec:linear} is constructed explicitly, so its
existence needs no Ext calculation. Here we show that it is the only
choice: in its weight it is unique up to scale,
\begin{equation}
	\begin{gathered}
		\Hom_\OY(k,M_\Phi)=\Ext^1_\OY(k,M_\Phi)=0,\\
		\Ext^2_\OY(k,M_\Phi)_{-2}\cong\C.
	\end{gathered}\label{eq:ext}
\end{equation}
This is used in the proof of Theorem~\ref{thm:dual}.

Here $\Ext^2_\OY(k,M_\Phi)_{-2}$ denotes the weight component of morphisms
$k\to M_\Phi(-2)[2]$, and $p_0$ is the generator of $M_\Phi(-2)$. It
carries the quantum numbers opposite to those of $\bD_+\bD_-$: spin $0$,
(vector, axial) R-charges $(-2,0)$ and dimension $-1$.

Neither a module map nor a class of degree one exists:
$\Hom_\OY(k,M_\Phi)=\Ext^1_\OY(k,M_\Phi)=0$.\footnote{A homomorphism from
	$k$ lands in the elements annihilated by every ghost, and in
	$M_\Phi=\C[\lbp,\lbm]$ the surviving ghosts annihilate no nonzero element.
	A $1$-cochain assigns $\gamma_i\in M_\Phi$ to $\xi_i$; the syzygy
	$K_{\bpl\bmi}$ gives $\lbp \gamma_{\bmi}=\lbm \gamma_{\bpl}$, so $\gamma_{\bpl}=\lbp a$
	and $\gamma_{\bmi}=\lbm a$ for some $a\in M_\Phi$, while $K_{+\bpl}$ and
	$K_{-\bmi}$ give $\gamma_\pm=0$; the cocycle is then the coboundary of $a$.}
In degree two, a cochain of weight zero assigns to each of the eight
generators $g$ of $\Omega^2$ a number $\varphi_g$, the coefficient of
$p_0$, and it is a cocycle if it respects the twelve relations among them.
There are no coboundaries, since $M_\Phi(-2)$ has nothing in the
weight needed to receive $\OY(-1)^4$. On $M_\Phi(-2)$ the unbarred
ghosts act by zero and $\lbp p_0$, $\lbm p_0$ are linearly independent, so
a relation forces every value that it multiplies by a barred ghost to
vanish:
\begin{itemize}
	\item the Koszul relations
	      $\lambda_iK_{jk}-\lambda_jK_{ik}+\lambda_kK_{ij}=0$ of the triples
	      $(+,\bpl,\bmi)$, $(\bpl,-,\bmi)$ and $(+,\bpl,-)$ give
	      $\lbm\varphi_{+\bpl}=\lbp\varphi_{+\bmi}$,
	      $\lbm\varphi_{\bpl-}=-\lbp\varphi_{-\bmi}$ and $\lbp\varphi_{+-}=0$, which
	      force these five values to vanish;
	\item the relations $\lbp(K_{+\bpl}+u_{\pplus})=\lbm(K_{-\bmi}+u_{\mminus})=0$
	      force $\varphi_{u_{\pplus}}$ and $\varphi_{u_{\mminus}}$ to vanish;
	\item $\varphi_{\bpl\bmi}$ is unconstrained, because in every relation
	      $K_{\bpl\bmi}$ is multiplied by $\lp$ or $\lm$.
\end{itemize}
This proves \eqref{eq:ext}; the group is
spanned by the cocycle $K^\vee_{\bpl\bmi}$ with
$K^\vee_{\bpl\bmi}(K_{\bpl\bmi})=p_0$ and zero on the other seven
generators. Since $c\neq0$ by Proposition~\ref{prop:yoneda}, it is a nonzero
multiple of $K^\vee_{\bpl\bmi}$; lifting the identity of $k$ along the splice
\eqref{eq:linear-splice} shows that in fact $c=-K^\vee_{\bpl\bmi}$. The
gluing thus pairs the chiral superfield with $K_{\bpl\bmi}$, the syzygy
relating the two dropped ghosts. The sign plays no role for the cone, since
$\Cone(-c)\cong\Cone(c)$.

\subsection{A finite CE model}\label{app:finite-model}
The class $c$ is a morphism in the derived category, and the analysis
above is carried out over $\OY$. Here we record that it is also realized by
a strict $\CE{2}$-module, on which the translation ghosts act. The model
has twelve free generators over the polynomial ring $R$ and is obtained by
deleting one exterior pair from the Koszul complex. Start with
\[
	K_R=R\otimes\Lambda(\xi_+,\xi_{\bpl},\xi_-,\xi_{\bmi}),
	\qquad d_K\xi_i=\lambda_i,
\]
and let $L_R$ be spanned by monomials not containing
$\xi_{\bpl}\xi_{\bmi}$. Since $d_K$ only removes exterior generators,
this is a subcomplex, with free $R$-ranks $(1,4,5,2)$. The quotient is the Koszul complex
of $(\lp,\lm)$, shifted by the deleted barred pair:
\[
	K_R/L_R=R\,\xi_{\bpl}\xi_{\bmi}\otimes\Lambda(\xi_+,\xi_-)
	\simeq M_\Phi(-2)[2].
\]
Equip these complexes with translation-ghost actions
\begin{equation}
	\vpp\mapsto\lbp\xi_+\wedge,\qquad
	\vmm\mapsto\lbm\xi_-\wedge.\label{eq:ce-actions}
\end{equation}
They preserve $L_R$, square to zero, anticommute, and satisfy
$d_Kv_{\pmpm}+v_{\pmpm}d_K=\lambda_\pm\bar\lambda_\pm$.
In the CE algebra the torsion quadrics are exact, not zero, and these
actions supply the required homotopies. The augmentations
$K_R\to k$ and $K_R/L_R\to M_\Phi(-2)[2]$ are CE-linear
quasi-isomorphisms, and the quotient map represents the class $c$ up to
sign. Indeed, a semifree resolution of $k$ over $\CE{2}$, mapped
$\CE{2}$-linearly to $K_R$, identifies the connecting class of
$0\to L_R\to K_R\to K_R/L_R\to0$ with the cochain $K^\vee_{\bpl\bmi}=-c$ of
Appendix~\ref{app:linear-class}. Hence
\begin{equation}
	L_R[1]\simeq\Cone\bigl(k\xrightarrow{c}M_\Phi(-2)[2]\bigr),
	\qquad L_R\simeq M_\Sigma.\label{eq:finite-model}
\end{equation}
With the specified CE action, reconstruction gives the complex-linear
multiplet with differential
\[
	d_K+\mathcal Q-2i\bigl(\lbp\xi_+\wedge\partial_{\pplus}
	+\lbm\xi_-\wedge\partial_{\mminus}\bigr).
\]
The last term comes from the translation-ghost actions
\eqref{eq:ce-actions}; it is invisible in any model over $\OY$, where the
quadrics vanish.

\paragraph{Component fields.}
The model also gives the off-shell content directly. As an $R$-complex,
$L_R$ is free and minimal: every entry of $d_K$ is a ghost. At zero
momentum the translation terms drop out, so by the discussion of component
fields in Section~\ref{sec:conventions} the components of $\A(L_R)$
correspond to the generators of $L_R$. A generator
$\xi_{i_1}\cdots\xi_{i_p}$ has resolution degree $-p$ and weight $p$, hence
ghost number zero, and parity $p$ modulo two. The ranks $(1,4,5,2)$ thus
give $1+5=6$ even and $4+2=6$ odd components, the $6+6$ of complex-linear
matter. Equivalently, the full Koszul complex $K_R$, with ranks
$(1,4,6,4,1)$, gives the $8+8$ components of an unconstrained superfield, and
the deleted part $\xi_{\bpl}\xi_{\bmi}\,\Lambda(\xi_+,\xi_-)$, with ranks
$(1,2,1)$, removes the $2+2$ components of the chiral target of
$\bD_+\bD_-$.

For the twisted complex-linear and conjugate multiplets the finite model
deletes the monomials containing $\xi_{\bpl}\xi_-$, resp.\ $\xi_+\xi_-$,
with the homotopies exchanged accordingly.

\section{Betti numbers of the gauge rings}\label{app:betti}
The Betti numbers of the ordinary, twisted and semichiral vector rings,
and of the LVM module, follow from their evaluation-at-the-origin sequences.
Let $R_1$ be the span of the four ghosts, and for a coordinate subspace
$Z_a$ let $W_a\subset R_1$ be the span of the ghosts vanishing on $Z_a$, so
that $\C[Z_a]=R/(W_a)$ has the Koszul resolution on $W_a$ and
$\Tor_i^R(\C[Z_a],\C)=\Lambda^iW_a$ in internal degree $i$.

\begin{lemma}\label{lem:betti}
	Let $0\to M\to\bigoplus_{a=1}^n\C[Z_a]\to k^n/k\to0$ be a restriction
	sequence. Then $\Tor_i^R(M,\C)_j=0$ unless $j\in\{i,i+1\}$, and
	\[
		\Tor_i(M)_i=\ker\beta_i,\qquad\Tor_i(M)_{i+1}=\operatorname{coker}\beta_{i+1},
		\qquad
		\beta_i:\bigoplus_a\Lambda^iW_a\to(k^n/k)\otimes\Lambda^iR_1,
	\]
	where $\beta_i$ is induced by the inclusions $W_a\subset R_1$. If the $W_a$ have
	zero common intersection, $\beta_i$ is injective for $i\ge1$, and $\beta_0$ is
	onto with one-dimensional kernel.
\end{lemma}
\begin{proof}
	In the long exact $\Tor$ sequence,
	$\Tor_i(\bigoplus_a\C[Z_a])$ lives in internal degree $i$ and
	$\Tor_i(k^n/k)=(k^n/k)\otimes\Lambda^iR_1$ also in degree $i$, so
	$\Tor_i(M)_j$ sits between $\Tor_{i+1}(k^n/k)_j$ and
	$\Tor_i(\bigoplus\C[Z_a])_j$, which vanish unless $j=i+1$ or $j=i$
	respectively. The map on $\Tor$ induced by $\C[Z_a]\to k$ is the inclusion
	$\Lambda^iW_a\subset\Lambda^iR_1$. An element of the kernel of $\beta_i$ is a
	tuple $(w_a)$ with all $w_a$ equal to a common $w\in\bigcap_a\Lambda^iW_a$,
	which for coordinate subspaces is $\Lambda^i(\bigcap_aW_a)$.
\end{proof}
With $\dim\Lambda^iR_1=\binom4i$ this gives:
\begin{center}\small
	\begin{tabular}{lllll}\toprule
		ring                                 & pieces      & $\dim W_a$ & ranks of $\beta_0,\dots,\beta_4$      & Betti numbers   \\\midrule
		$M_V$, $M_{\widetilde V}$, $\OYd{4}$ & two planes  & $2$        & $1,4,2,0,0$ (target $1,4,6,4,1$)      & $(1,4,4,1)$     \\
		$M_{\mathbb V}$                      & four axes   & $3$        & $3,12,12,4,0$ (target $3,12,18,12,3$) & $(1,6,8,3)$     \\
		$M_{\mathcal V}$                     & four planes & $2$        & $3,8,4,0,0$ (target $3,12,18,12,3$)   & $(1+4,14,12,3)$ \\\bottomrule
	\end{tabular}\end{center}

\section*{Acknowledgments}
We thank Rikard von Unge and Johannes Walcher for helpful discussions.

\end{document}